\documentclass[twocolumn,showpacs,amsmath,amssymb,iopart,floatfix]{revtex4-1}
\usepackage[latin1]{inputenc}
\usepackage[american]{babel}
\usepackage[T1]{fontenc}
\usepackage{mathrsfs}
\usepackage{hyperref}
\usepackage{multirow}
\usepackage{bm}
\usepackage{epsfig}
\usepackage{graphicx}
\usepackage{subeqnarray}
\usepackage{bbold}
\usepackage{upgreek}
\usepackage{url}

\hypersetup{colorlinks=true, urlcolor=blue, citecolor=blue, filecolor=blue, linkcolor=blue}

\begin{document}

\title{X-ray fluorescence spectrum of highly charged Fe ions driven by strong free-electron-laser fields}

\date{\today}

\author{Natalia~S.~Oreshkina${}^*$}
\affiliation{Max-Planck-Institut f\"{u}r Kernphysik, Saupfercheckweg 1, 69117 Heidelberg, Germany}
\email{Natalia.Oreshkina@mpi-hd.mpg.de}
\author{Stefano~M.~Cavaletto}
\affiliation{Max-Planck-Institut f\"{u}r Kernphysik, Saupfercheckweg 1, 69117 Heidelberg, Germany}
\author{Christoph~H.~Keitel}
\affiliation{Max-Planck-Institut f\"{u}r Kernphysik, Saupfercheckweg 1, 69117 Heidelberg, Germany}
\author{Zolt\'{a}n~Harman}
\affiliation{Max-Planck-Institut f\"{u}r Kernphysik, Saupfercheckweg 1, 69117 Heidelberg, Germany}

\begin{abstract}

The influence of nonlinear dynamical effects is analyzed on the observed spectra of controversial 3C and 3D astrophysically relevant x-ray lines in neonlike Fe${}^{16+}$ and the A, B, C lines in natriumlike Fe${}^{15+}$ ions. 
First, a large-scale configuration-interaction calculation of oscillator strengths is performed with the inclusion of higher-order electron-correlation effects. 
Also, quantum-electrodynamic corrections to the transition energies are calculated. 
Further considered  dynamical effects provide a possible resolution of the discrepancy between theory and experiment found by recent x-ray free-electron-laser measurements of these controversial lines. 
We find that, for strong x-ray sources, the modeling of the spectral lines by a peak with an area proportional to the oscillator strength is not sufficient and nonlinear dynamical effects have to be taken into account. 
Thus, we advocate the use of light-matter-interaction models also valid for strong light fields in the analysis and interpretation of the associated astrophysical and laboratory spectra. 
We investigate line-strength ratios distinguishing between the coherent and incoherent parts of the emission spectrum. 
In addition, the spectrum of Fe${}^{15+}$, an autoionizing ion which was also present in the recent laboratory experiment, is also analized.

\end{abstract}

%
%

\maketitle

\section{Introduction}

During the last decades, highly charged ions have become an object of interest in a wide range of physics. 
Well-developed theoretical models allow for highly accurate predictions, which, together with modern experimental techniques, enable the  benchmark of atomic properties, tests of quantum electrodynamics in strong electric and magnetic fields~\cite {PhysRevLett.95.233003, PhysRevLett.100.073201, Shabaev2006109}, the determination~\cite{PhysRevLett.96.253002, sturm_nature_2014} and the search for possible variations~\cite{PhysRevLett.105.120801, PhysRevLett.113.030801, PhysRevLett.94.243002, PhysRevLett.114.150801, Bekker2015} of the fundamental constants. 
In astrophysical research, atomic data of highly charged ions, e.g., transition energies and probabilities, oscillator strengths, collisional and recombination cross sections, together with astrophysically measured spectra, allow one to determine the temperature, density, velocity, and element composition of various distant objects such as stars, x-ray binaries, black hole accretion discs, or active galactic nuclei~\cite{Simon2009, Simon2010, Schnorr2013, Steinbruegge2015}.

The x-ray emission lines of highly charged Fe ions are among the brightest in astrophysical spectra~\cite{Paerels2003}. 
Several observations were performed with the space laboratories Chandra and XMM-Newton (see, e.g., \cite{Huenemoerder2011,Paerels2003,Xu2002,Mewe2001}).
Particular interest has been focused on two strong $2p\to 3d$ E1 transitions in Fe${}^{16+}$, customarily denoted as 3C and 3D, at 826 and 812~eV, respectively.
For both lines the initial state is spherically symmetric, with a total angular momentum $J=0$, and excited states have the same symmetry, $J=1$. 
The line-strength ratio of those two lines was observed to disagree with theoretical predictions~\cite{Bhatia1992, Cornille1994, Safronova2001, Chen2001}.
In an astrophysical setting, this disagreement was considered for a long time to originate from plasma effects and from the co-existence of different charge states of Fe ions; and in laboratory measurements, the unavoidable electron-impact excitation of the ions was considered a possible cause of the discrepancy.
To exclude those effects, a laser spectroscopic experiment in the x-ray regime~\cite{Bernitt2012} was recently carried out at an x-ray free-electron-laser (XFEL) facility \cite{Emma2010}. 
As a result of this experiment, a disagreement between all state-of-the-art theoretical predictions (ratio of the 3C and 3D oscillator strength around 3.5 and above) and an experimental line-strength ratio of 2.61(23) was stated~\cite{Bernitt2012}, which could be explained either with incorrect or insufficient knowledge of many-body atomic-structure theory, or with the presence of additional unknown effects.

In a recent Letter \cite{Oreshkina_prl_2014}, motivated by the above discrepancy, we reported on a possible mechanism to reduce the current disagreement between theoretical predictions and experimental results \cite{Bernitt2012}. 
Similar results were later found by another theoretical study \cite{Loch_2015}. 
Here, we present a more complete theory of x-ray-ion interactions by calculating higher-order electron-correlation and quantum-electrodynamic (QED) corrections, along with nonlinear dynamical effects contributing to the 3C/3D line-strength ratio. 
We also present calculations on the coherent and incoherent part of the resonance fluorescence spectrum, whereby a method for checking the limitations of the applicability of the weak-field atomic theory is proposed.
Finally, we analyze the presence of autoionizing Fe${}^{15+}$ ions in the trap and show that a wide XFEL intensity range should be considered. 

The paper is structured as follows: in Subsection \ref{Higher-order correlation correction to the oscillator strengths} we discuss the many-body atomic-structure calculations for the  energies and Einstein coefficients of the transitions under consideration including higher-order contributions. 
Screened QED corrections to the transition energies are presented in Subsection \ref{Screened quantum electrodynamic corrections to the transition energies}. 
In Section \ref{Strong-field effects} we present a theoretical model for nonlinear dynamical effects (\ref{Modeling of ion--x-ray interaction}, \ref{Line-strength-ratio calculation}), results of our simulations compared with current experimental data \cite{Bernitt2012} (\ref{Results and comparison with experimental line-strength ratios}), and possible additionally contributing effects (\ref{Influence of the atomic calculations uncertainty}, \ref{Influence of Fe15+ lines}).

\section{Improvement of atomic-structure parameters}
\label{Improvement of atomic-structure parameters}
As a first step towards resolving the discrepancy between theoretical predictions and experimental results, atomic-structure parameters, such as transition energies and Einstein coefficients, need to be refined. 
The calculation of the transition energies and Einstein coefficients with the configuration-interaction Dirac-Fock-Sturm (CI-DFS) method is considered in Subsection \ref{Higher-order correlation correction to the oscillator strengths}. 
Additionally, QED corrections to the transition energies are taken into account in an {\it ab initio} manner in Subsection~\ref{Screened quantum electrodynamic corrections to the transition energies}.

\subsection{Higher-order correlation electron-correction to the oscillator strengths}
\label{Higher-order correlation correction to the oscillator strengths}

In our calculations, the CI-DFS method~\cite{tup2003, Tupitsyn2005, SoriaOrts2006} was used for atomic-structure-theory calculations of the transition energies and dipole-transition rates in a large-scale setting. 
In the Breit approximation, the Hamiltonian of the relativistic $N_e$-electron system is given as the sum of the single-particle Dirac Hamiltonians $h^D_i$ and the interelectronic-interaction operator $u_{ij}$, which, with the usual notations~\cite{tup2003}, can be written as
\begin{align}
H &=\sum_i^{N_e} h^D_i + \frac{1}{2}\sum_{i \neq j} u_{ij}, \\
h^D &= c (\vec{\alpha} \cdot \vec{p} ) + (\beta -1)mc^2 + V_{\rm nucl}(r), \\
u_{ij} &=  {\alpha\hbar c}  \left[ \frac{1}{r_{ij}} -
\frac{\vec{\alpha}_i\cdot\vec{\alpha}_j}{2r_{ij}} - \frac{(\vec{\alpha}_i\cdot
\vec{r}_{ij}) (\vec{\alpha}_j\cdot \vec{r}_{ij})}{2r_{ij}^3} \right].
\end{align}

The numerical solution of the Dirac-Fock equation gives one-electron radial wave functions for the occupied orbitals,
while solutions of the generalized Dirac-Fock-Sturm equation are used as virtual orbitals with positive and negative energies.
Configuration state functions are built from $jj$-coupled Slater determinants with such one-electron wave functions. 
The total many-electron wave function $| \Psi_J \rangle$ with a total angular momentum $J$ is given as a linear combination of a large number $N_c$ of configuration state functions
$| \Phi^\varepsilon_J \rangle$:
\begin{eqnarray}
|\Psi_J \rangle = \sum_{\varepsilon=1}^{N_c} c_\varepsilon |\Phi_J^\varepsilon \rangle ,
\end{eqnarray}
with the configuration coefficients $c_\varepsilon$.
The Einstein coefficients $A_{eg}$ and oscillator strengths $f_{eg}$ are calculated with such many-electron wave
functions representing the $2p^6$ ground $(g)$ and $2p^53d$ excited $(e)$ 
states of the ion~\cite{Grant1974}:
\begin{eqnarray}
A_{eg} &=& \frac{4 \pi^2 e^2 c^2}{(2J_{e}+1)\omega_{eg}} \sum_{M_{e},M_{g}}\sum_{\vec{k}/k, \sigma}
\left| \langle {e} | \vec{\alpha}\vec{\epsilon}_{\vec{k}\sigma}e^{-i\vec{k}\vec{r}}| {g} \rangle \right|^2\,, \nonumber \\
f_{eg} &=& \frac{2J_e+1}{2J_g+1}\frac{A_{eg}mc^3}{2\omega^2_{eg}e^2} \,.
\end{eqnarray}
Here, $c$, $e$ and $m$ stand for the speed of light, the elementary charge, and the electron mass,
respectively, and $\vec{\alpha}$ and $\vec{\epsilon}_{\vec{k}\sigma}$ stand for the vector of Dirac alpha matrices and the photon polarization unit vector. 
The summation goes over the magnetic quantum numbers of the initial state $M_i$ and final state $M_f$, over the polarization $\sigma=1,2$ of the emitted photon, and the integration goes over the direction $\vec{k}/k$ of the emitted photon.
The line strength is  given as the integrated peak area of the resonance fluorescence cross section~\cite{Grant1974,Foot}:
\begin{eqnarray}
S_{eg} = \frac{\pi^2c^2\hbar^3}{(\hbar\omega)^2}\frac{g_e}{g_g}A_{eg} \propto f_{eg}\,.
\end{eqnarray}
Thus for low driving-field intensities the oscillator strength is proportional to the line strength.

\begin{table}
\begin{tabular}{l l c r}
\hline\hline
Line	& Transition				& Energy, eV	& \quad $\Gamma_{eg}$, $1/$s \\
\hline
3D	& $(2p^6)_{0} \rightarrow ((2p^5)_{3/2}3d_{5/2})_{1}$	& 812	& $6.02\cdot10^{13}$ \\
3C	& $(2p^6)_{0} \rightarrow ((2p^5)_{1/2}3d_{3/2})_{1}$	& 826	& $2.21\cdot10^{14}$ \\
\hline\hline
\end{tabular}
\caption{Transitions of Fe${}^{16+}$, with their notations, energies, and spontaneous-decay rates $\Gamma_{eg}$ calculated with the CI-DFS method. } \label{tab:fe16}
\end{table}

To build the configuration space, the restricted-active-space~\cite{Malmqvist2002230} approach was used.
In our calculations, all orbitals up to $3d$ form the occupied shells' space, orbitals from $4d$ to $5p$ belong to the active space, and orbitals from $4s$ form the open shells' space.  
Single and double excitation from the occupied shells' space and from the open shells' space as well as triple excitations within the active space were included into the calculations, leading to a huge number of configurations of the order of $N_c=100\, 000$.
The results of our CI-DFS calculations are presented in Table~\ref{tab:fe16}.
To estimate the accuracy of our results, we performed a series of calculations with the one-electron functions up to $n=6, 7$, or 8, 
where $n$ is the principal quantum number. 
Thereby, our calculated oscillator-strength ratio converged with three-digit precision.
As a result, for the case of single and double excitations included, the 3C/3D oscillator-strength ratio is 3.57, while the
contribution of the triple excitation is as low as -0.01. Thus, our results confirm earlier theoretical calculations and disprove
the significance of remaining higher-order correlation effects such as triple excitations, suggesting that the discrepancy of
theory and experiment is not funded in the inaccurate description of the ions' electronic structure.

\subsection{Screened quantum-electrodynamic corrections to the transition energies}
\label{Screened quantum electrodynamic corrections to the transition energies}

In order to match the accuracy of the experimental transition energies $\omega_{eg} = \omega_e-\omega_g$, additional QED corrections were taken into account in an {\it ab initio} manner. 
The QED corrections in first order in the fine-structure constant $\alpha$ consist of the self-energy (SE) and vacuum-polarization (VP) terms.
The SE correction to the single-electron state $|a\rangle$ with energy $\varepsilon_a$ is given by the expression 
\begin{align}\label{se}
  \Delta E^{(a)}_{\rm SE} & ={\alpha\hbar c }\frac{i}{2\pi}\int\limits_{-\infty}^{\infty} d\omega \sum_n
    \frac{\langle an|(\alpha^\mu \alpha^\nu D_{\mu\nu}(\omega,x_1-x_2)|na\rangle}
    {\varepsilon_a - \omega - \varepsilon_n(1-i0)}  \nonumber \\ 
     &- \langle a|\gamma^0\delta m{c^2}|a\rangle,
\end{align}
where $\omega$ is the variable frequency of the emitted and re-absorbed virtual photon, $D_{\mu\nu}$ is the photon propagator with $\mu$ and $\nu$ being Minkowski indexes, $\alpha^\mu$ and  $\gamma^0$ are Dirac matrices, $\delta m$ is the mass counter-term, and the summation
goes over the complete Dirac spectrum, including the positive and negative continua.
To separate the ultraviolet divergences, Eq. \eqref{se} is decomposed into zero-, one-,
and many-potential terms following the procedure from Ref.~\cite{Yer1999}. The zero-potential and one-potential terms are 
calculated in momentum space employing numerical Fourier transforms of bound-state wave functions.
The residual part of the SE correction, the so-called many-potential term, is calculated in coordinate space. 
The angular integrations and the summation over the intermediate angular-momentum projections are carried out in a standard algebraic manner. 
The many-potential term involves an infinite summation over the relativistic angular quantum number $\kappa = \pm(j+1/2)$. The summation is terminated at a maximum value
$|\kappa|=15-20$, while the residual part of the sum is approximated by a least-square inverse-polynomial fit.
For any given $\kappa$, the summation over the Dirac spectrum is performed using the dual-kinetic-balance approach~\cite{dkb}
with basis functions constructed from B-splines, taking into account the finite distribution of the nuclear charge.
The VP correction was calculated in the Uehling approximation~\cite{uehling, Mohr1998} with the potential
\begin{align}
 U_{\rm VP, Ue}(r) &= -\alpha Z \frac{\alpha {\hbar c}}{3\pi}\int_1^\infty dt \sqrt{t^2 -1}\biggl(\frac{2}{t^2}+\frac{1}{t^4}\biggr) \nonumber \\
 & \times \int d^3 {\bf r}\exp({-2|{\bf r} - {\bf r'}|{mc/\hbar}})\frac{\rho_{\rm nucl}({\bf r'})}{|{\bf r} - {\bf r'}|}.
\end{align}
%

In this work, the interelectronic-interaction effects to the QED correction were approximated by evaluating the single-electron QED terms with wave functions belonging to an effective potential accounting for the screening of the remaining 9 electrons. We used different screening potentials for the assessment of the uncertainty:
the core-Hartree (CH) potential
\begin{align*}
V_{\rm CH,a}(r) = \alpha {\hbar c}\int_0^\infty dr' \frac{1}{\max(r,r')}\rho_a(r')\,,
\end{align*}
the Kohn-Sham (KS) potential
\begin{align}\label{eq:KS}
V_{\rm KS}(r) &= \alpha{\hbar c} \int_0^\infty dr' \frac{1}{\max(r,r')}\rho_t(r')  \\
     &- \frac{2}{3}\frac{\alpha {\hbar c}}{r} \biggl[ \frac{81}{32\pi^2} r\rho_t(r)\biggr]^{1/3} 
     -\frac{\alpha {\hbar c}}{r}\left[1-{\rm e}^{-(Ar)^2}\right]\,, \nonumber
\end{align}
and the Dirac-Slater (DS) potential
\begin{align}\label{eq:DS}
V_{\rm DS}(r) &= \alpha {\hbar c}\int_0^\infty dr' \frac{1}{\max(r,r')}\rho_t(r') \\
    &- \frac{\alpha{\hbar c}}{r} \biggl[ \frac{81}{32\pi^2} r\rho_t(r)\biggr]^{1/3} \nonumber 
    -\frac{\alpha{\hbar c}}{r}\left[1-{\rm e}^{-(Ar)^2}\right]\,.
\end{align}
Here, $\rho_a$ is the density of all one-electron orbitals excluding that of the electron participating in the transition,  labeled by $a$:
\begin{equation}
\rho_a(r)={\frac{mc}{\hbar}}\sum_{b\neq a}q_b \left[ G_b^2(r)+F_b^2(r) \right] \,,
\end{equation}
with non-integer occupation numbers $q_b$ taken from the results of the CI-DFS calculation. Analogously, $\rho_t$ is the total electron density
\begin{equation}
\rho_t(r)={\frac{mc}{\hbar}}\sum_{b}q_b \left[ G_b^2(r)+F_b^2(r) \right]\,.
\end{equation}
The last term in the KS and DS potentials stands for the Latter correction~\cite{Latter} to provide a proper asymptotic behavior at large radial distances.
The parameter $A$ in Eqs.~\eqref{eq:KS} and \eqref{eq:DS} was chosen here to be $\alpha Z/10$, found by numerical optimization.
The CH potential was calculated in one step, while the KS and DS potentials were evaluated in a self-consistent manner. 
The total QED result is determined as a sum of the one-electron contributions weighted by the occupation numbers $q_a$, taken from the CI-DFS calculations:
\begin{equation}
\Delta E_{\rm QED} = \sum_aq_a\Delta E^{(a)}_{\rm QED}.
\end{equation}
In Table \ref{tab:QED} results of the calculations in the independent-particle approximation (i.e., without the inclusion of interelectronic-interaction effects by the screening potential), and for CH, KS and DS screening potentials are presented.
Our total QED correction result is defined as the average of the results with three screening potentials.
The difference between the results for the QED corrections with various screening potentials can be considered as an approximate uncertainty.
As one can see from  Table~\ref{tab:QED}, screened QED corrections to the transition energies for both 3C and 3D lines $|\Delta E_{\rm QED}|$ are less than 0.03~eV, therefore, with the transition energies on a level of 800~eV, they can be safely neglected.
\begin{table}
\begin{center}
\begin{tabular}{lrrrrr}
\hline\hline
Line	& Unscreened 	& CH	& KS	& DS	& Total QED 	\\
\hline
3C	& -0.0033	&  0.0180	&  0.0336	&  0.0330	&  0.0282 \\
3D	& -0.0357	& -0.0241	& -0.0042	& -0.0058	& -0.0113 \\
\hline\hline
\end{tabular}
\caption{
QED correction to the 3C and 3D level energies in Fe${}^{16+}$ without screening, and with various screening potentials (CH, KS, and DS),
in units of eV.\label{tab:QED}}
\end{center}
\end{table}

\section{Strong-field effects}
\label{Strong-field effects}

In the present Section, nonlinear dynamical effects of light-matter interaction are investigated. By showing how these may have played 
an important role in the experiment of Ref.~\cite{Bernitt2012}, we provide a possible explanation of the discrepancy with theory reported therein.

The general theoretical model is introduced in \ref{Modeling of ion--x-ray interaction}, while in \ref{Line-strength-ratio calculation} we show how to calculate the line-strength ratios measured in Ref.~\cite{Bernitt2012}, distinguishing also between coherent and incoherent contributions to the total detected electromagnetic energy. Theoretical predictions for Fe${}^{16+}$ 3C and 3D lines are presented in~\ref{Results and comparison with experimental line-strength ratios} and compared with the experimental results of Ref.~\cite{Bernitt2012}. 
While in \ref{Influence of the atomic calculations uncertainty} we discuss the influence of atomic-calculation uncertainties, in \ref{Influence of Fe15+ lines} we conclude this Section by considering strong-field effects to the strengths of Fe${}^{15+}$ lines, which in Ref.~\cite{Bernitt2012} were also measured and subsequently disentangled from the reported Fe${}^{16+}$ lines.

\subsection{Modeling of ion--x-ray interaction}
\label{Modeling of ion--x-ray interaction}

In order to account for nonlinear dynamical effects dependent upon the intensity of the x-ray pulses, we model the time evolution of the driven ions via a master-equation approach. 
The atomic system is therefore described as a two-level system with a ground state $|g\rangle$ and an excited state $|e\rangle$, with a transition 
energy $\omega_{eg}$. 
This is used to model {the x-ray transitions responsible for} the 3C and 3D lines in Fe${}^{16+}$, as well as the A, B, and C lines in Fe${}^{15+}$. 
In both ionic systems, the excited states are sufficiently separated, with transition energies differing by more than the x-ray-pulse bandwidth and the natural linewidths, such that we are allowed to treat each transition as an independent two-level system. 
{More details on the validity of this two-level approximation are provided in Appendix~\ref{Two-level model}.}

The ionic system is described via the density matrix $\hat{\rho}(t)$, of elements $\rho_{ij} = \langle i |\hat{\rho}(t)|j\rangle$, $i,\,j\in\{g,\,e\}$, 
evolving in time under the master equation \cite{Scully,Kiffner2010}
\begin{equation}
\frac{d\hat{\rho}}{dt} = - \frac{i}{\hbar}[\hat{H},\hat{\rho}(t)] + \mathcal{L}\hat{\rho}(t) +\mathcal{D}\hat{\rho}(t).
\label{eq:mastereq}
\end{equation}
The first term $- \cfrac{i}{\hbar}\,[\hat{H},\hat{\rho}(t)]$ models the coherent evolution of the two-level system interacting with an external electric field. 
The total Hamiltonian 
\begin{equation}
\hat{H} = \hat{H}_0 + \hat{H}_{\mathrm{int}}
\end{equation}
is the sum of the electronic-structure Hamiltonian 
\begin{equation}
\hat{H}_0 = \sum_{i \in \{g,e\}}\hbar\omega_i |i\rangle\langle i|
\end{equation}
and of the interaction Hamiltonian 
\begin{equation}
\hat{H}_{\mathrm{int}} = -\frac{\hbar\Omega_R(t)}{2}\,e^{i\omega_{\mathrm{X}}t}|g\rangle\langle e| -\frac{\hbar\Omega_R^*(t)}{2}\,e^{-i\omega_{\mathrm{X}}t}|e\rangle\langle g|, 
\label{eq:interactionhamiltonian}
\end{equation}
describing, in the rotating-wave approximation, the interaction of the two-level system with an external time-dependent 
electric field $\mathscr{E}(t)=E(t)\cos(\omega_{\mathrm{X}}t+\psi(t))$ of x-ray carrier frequency
$\omega_{\mathrm{X}}$, envelope $E(t)$, and phase $\psi(t)$. In Eq.~(\ref{eq:interactionhamiltonian}), we introduced the complex time-dependent Rabi frequency 
\begin{equation}
\Omega_R(t)=\frac{e}{\hbar}E(t)\langle g| \hat{z} | e \rangle e^{i\psi(t)},
\end{equation}
which is proportional to the {dipole-moment matrix element $\langle g| e\hat{z} | e \rangle$ and to the} square root of the x-ray intensity
\begin{equation}
I(t)  = \frac{1}{8\pi\alpha} \frac{e^2}{\hbar} E^2(t). 
\end{equation}
{Here, $\hat{z}$ is the position operator along the quantization axis.}

The Lindblad superoperator $\mathcal{L}\hat{\rho}(t)$ models the norm-conserving spontaneous decay from the excited state $|e\rangle$ to 
the ground state $|g\rangle$ at a decay rate $\Gamma_{eg} = \hbar A_{eg}$. 
While both excited states in Fe${}^{16+}$ are nonautoionizing, Auger decay need be taken into account for the three excited states of Fe${}^{15+}$ modeled here. 
This is included in the norm-nonconserving term $\mathcal{D}\hat{\rho}(t)$, which would not be present in the Lindblad form of the master equation. 
Auger decay destroys the two-level system by further ionizion of Fe${}^{15+}$ to levels which, however, do not have to be accounted for explicitly. 
Here, we model this autoionization process as a loss of population and coherence from the excited state $|e\rangle$ at the Auger-decay rate $\Gamma_{\mathrm{A}}$. 
The total decay rate $\Gamma_{\rm tot} = \Gamma_{eg} + \Gamma_{\mathrm{A}}$ is hence given by the sum of the spontaneous- and Auger-decay rates. 

We introduce the vector $\vec{R}(t)$ of the slowly varying components of the density matrix,
\begin{equation}
\vec{R}(t) = (\rho_{gg}(t),\,\rho_{ge}(t)\,e^{-i\omega_{\mathrm{X}}t},\,\rho_{eg}(t)\,e^{i\omega_{\mathrm{X}}t},\,\rho_{ee}(t))^{\mathrm{T}}\,,
\end{equation}
such that the master equation~(\ref{eq:mastereq}) can be written in the matrix form
\begin{equation}
\frac{d\vec{R}(t)}{dt} = {\bf M}(t)\vec{R}(t)\,,
\label{eq:master-equation}
\end{equation}
with the $4\times 4$ time-dependent matrix
\begin{eqnarray*}
{\bf M}(t) = \left(
\begin{array}{c c c c}
0 				& -i\frac{\Omega^*_R(t)}{2}		& i\frac{\Omega_R(t)}{2}		&  \Gamma_{eg}			\\
-i\frac{\Omega_R(t)}{2}		& i\Delta-\frac{\Gamma_{\rm tot}}{2}	& 0 					&  i\frac{\Omega_R(t)}{2}	\\
 i\frac{\Omega^*_R(t)}{2}	& 0 					& -i\Delta-\frac{\Gamma_{\rm tot}}{2}	& -i\frac{\Omega^*_R(t)}{2}	\\
0 				& i\frac{\Omega^*_R(t)}{2}		& -i\frac{\Omega_R(t)}{2}		& -\Gamma_{\rm tot}		\\
\end{array} \right),
\end{eqnarray*}
where $\Delta=\omega_{eg}-\omega_X$ denotes the detuning of the laser frequency from the transition energy. 

\subsection{Calculation of the line-strength ratio}
\label{Line-strength-ratio calculation}

In order to calculate the line-strength ratios reported in Ref.~\cite{Bernitt2012} and compare our theoretical predictions with the experimental results presented therein, we solve the system of differential equations~(\ref{eq:master-equation}) for time-dependent x-ray driving pulses, assuming natural initial conditions $\vec{R}_0=(1,0,0,0)^T$. 
{In the following we directly refer to the 3C and 3D transitions in Fe${}^{16+}$. Analogous formulas apply to the A, B, and C transitions in Fe${}^{15+}$ investigated in Subsection~\ref{Influence of Fe15+ lines}. 
Further details are provided in Appendix~\ref{Fe15+ line-strength ratios}.}

At fixed detuning $\Delta$, the total detected x-ray energy can be directly inferred from the time evolution of the two-level system. 
The total detected energy consists of two different contributions, which in the context of resonance fluorescence are usually referred to as coherent and incoherent energies. 
These two contributions exhibit different space-dependent features, i.e., their energies differently depend upon the position of the photon detector \cite{Jentschura20041, Scully, Kiffner2010}.

The electromagnetic energy due to the coherent part of the spectrum of resonance fluorescence is proportional to
\begin{equation}
\mathcal{E}_{\mathrm{coh}}(\Delta) \propto \Gamma_{eg} \omega_{eg} \int_{-\infty}^{\infty} |R_{eg}(t)|^2\,dt,
\label{eq:E-coh}
\end{equation}
where the time evolution of the coherence $R_{eg}(t)$ also depends on $\Delta$. 
In a linear sample with length $L$ larger than the other two dimensions, due to the superposition of the radiation emitted by $N$ different ions along the forward direction, i.e., along the propagation direction of the driving x-ray field, $\mathcal{E}_{\mathrm{coh}}$ features a high degree of directionality~\cite{PhysRevA.45.4706, PhysRevA.52.525}. 
As a result of these 
many-particle effects, inside an emission cone centered around the forward direction and with an opening angle 
$\phi = \sqrt{2\pi c/\omega_{eg}L}$,  $\mathcal{E}_{\mathrm{coh}}$ is proportional to $N^2$ and 
thus represents the main contribution to the total detected energy, 
$\mathcal{E} \approx \mathcal{E}_{\mathrm{coh}}$.

\begin{figure}[t]
\includegraphics[clip=true, width=.99 \columnwidth]{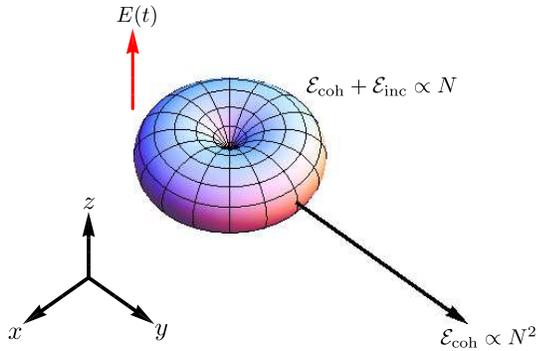}
\caption{(color online) Space-dependent distribution of the coherent and incoherent electromagnetic energies emitted by an oscillating dipole (such as a Fe${}^{16+}$ two-level system), driven by a $z$-polarized electric field propagating in the forward $y$ direction. The black arrow stands for the emission cone in the forward direction, along which the radiation emitted by separate ions sums up coherently.}
\label{fig:geometry}
\end{figure}

Out of this emission cone, however, the detected energy results from the independent superposition of the radiation emitted by $N$ 
two-level ions acting as oscillating dipoles. 
Both coherent and incoherent parts are proportional to $N$ and feature the angular distribution of an oscillating dipole, such that, along directions different from the forward direction, both contributions need be included (see Fig.~\ref{fig:geometry}). 
This total detected energy can be calculated from the atomic dynamics,
\begin{equation}
\mathcal{E}(\Delta) \propto \Gamma_{eg}\omega_{eg}\int_{-\infty}^{+\infty}R_{ee}(t)dt\,,
\label{eq:E-Delta}
\end{equation}
where the time-dependent population $R_{ee}(t)$ of the excited state depends on $\Delta$. 
Along directions different from the forward direction, where $\mathcal{E} = \mathcal{E}_{\mathrm{coh}} + \mathcal{E}_{\mathrm{inc}}$, it is therefore possible to calculate the contribution of the incoherent energy $\mathcal{E}_{\mathrm{inc}}(\Delta)$ via the difference of Eqs.~(\ref{eq:E-Delta}) and (\ref{eq:E-coh}).

For Fe${}^{16+}$ ions, the ratio of emitted energies is given by
\begin{equation}
S_{\rm 3C}/S_{\rm 3D} = \cfrac{\int d \Delta \mathcal{E}_{\rm 3C}(\Delta)}{\int d \Delta \mathcal{E}_{\rm 3D}(\Delta)}, \label{eq:spectrum}
\end{equation}
where $\mathcal{E}_{\rm 3C}(\Delta)$ and $\mathcal{E}_{\rm 3D}(\Delta)$ are the detected energies as a function of the detuning $\Delta$ from
the transition frequencies $\omega_{\rm 3C}$ and $\omega_{\rm 3D}$, respectively. In the forward direction, these are given by the coherent energies 
in Eq.~(\ref{eq:E-coh}), while outside of the emission cone in the forward direction they are obtained from the sum of coherent and incoherent 
energies given by Eq.~(\ref{eq:E-Delta}), see also Fig.~\ref{fig:geometry}. 

In Eqs.~(\ref{eq:E-coh}) and (\ref{eq:E-Delta}), the dependence upon the detection direction was not included explicitly. 
For Fe${}^{16+}$, however, this is not necessary when line-strength ratios are calculated. 
The excited states associated with the 3C and 3D lines feature identical symmetry (total angular momentum $J=1$), while the ground state is spherically symmetric ($J=0$). 
As a result, the angular distribution of the emitted radiation is the same for both transitions~\cite{Bernitt2012} and does not affect the ratio in Eq.~(\ref{eq:spectrum}).

Finally, in the case of Fe${}^{16+}$ ions, when no Auger decay is present, it is interesting to further investigate the limit of infinitely 
long coherent pulses. The state of a system driven by a continuous-wave driving field $E(t) = \bar{E}$, $\psi(t) =0$, $I(t) = \bar{I}$, with 
correspondingly constant Rabi frequency $\bar{\Omega}_R$, converges
for $t\rightarrow \infty$ to the stationary solution \cite{Scully,Foot}
\begin{equation}
\bar{R}_{ee}(\Delta) = \frac{\bar{\Omega}_R^2}{4\Delta^2 + \Gamma_{eg}^2 + 2\bar{\Omega}_R^2}\,.
\end{equation}
Far from the forward direction, where both coherent and incoherent energies are measured, the energy detected per unit time for a given detuning $\Delta$ is equal to
\begin{equation}
\mathcal{I}(\Delta) \propto \Gamma_{eg}\omega_{eg}\bar{R}_{ee}(\Delta)\,,
\end{equation}
yielding the intensity ratio
\begin{equation}
\begin{split}
S_{\rm 3C}/S_{\rm 3D} &= \frac{\int {\rm d}\Delta \mathcal{I}_{\rm 3C}(\Delta)}{\int {\rm d}\Delta \mathcal{I}_{\rm 3D}(\Delta)} \\
&= \frac{\Gamma_{\mathrm{\rm 3C}}\omega_{\mathrm{\rm 3D}}^2}{\Gamma_{\mathrm{\rm 3D}}\omega_{\mathrm{\rm 3C}}^2}\,\sqrt{\frac{1 + I/I_{\rm sat,3D}}{1+ I/I_{\rm sat, 3C}}},
\end{split}
\label{eq:analyticalratio}
\end{equation}
where for each line we have introduced the saturation intensity 
\begin{equation}
I_{\mathrm{sat}} = I \frac{\Gamma_{eg}^2}{2 \bar{\Omega}^2_R} = \frac{\hbar}{16\pi\alpha}\frac{\Gamma_{eg}^2}{|\langle g| \hat{z} | e \rangle|^2}
= \frac{\hbar}{c^2e^2}\frac{\Gamma_{eg}\omega_{eg}^3}{12 \pi}.
\end{equation}
For a weak exciting field ($I\ll I_{\rm sat}$), this agrees with the linear theory of resonance fluorescence, predicting a line-strength ratio of 3.56, equal to the ratio of the oscillator strengths. 
This weak-field limit was the implicit assumption of Ref.~\cite{Bernitt2012} and previous studies, where observed line-strength ratios were compared to the ratio of the oscillator strengths. 
However, for field intensities $I$ comparable to or larger than the saturation intensity $I_{\rm sat}$, Eq.~(\ref{eq:analyticalratio}) shows that nonlinear effects become important and the line-strength ratio converges to $S_{\rm 3C}/S_{\rm 3D} \rightarrow 7.03$, significantly different from the ratio of the oscillator strengths. 
For the Fe transitions studied here, with $I_{\rm sat} \approx$ $10^{11}$ W/cm$^2$, the intensity of LCLS pulses is typically on or above this order of magnitude. 
This justifies a thorough investigation of nonlinear dynamical effects. 

\subsection{Results and comparison with experimental line-strength ratios}
\label{Results and comparison with experimental line-strength ratios}

In this Subsection we present theoretical results for a range of XFEL pulse intensities, durations, and bandwidths. 
Since in the experiment of Ref.~\cite{Bernitt2012} the detected electromagnetic energy was measured out of the forward direction, we will first focus on this particular setup, studying the ratio~(\ref{eq:spectrum}) of the line strengths given by Eq.~(\ref{eq:E-Delta}). 
We will subsequently discuss the different contributions owing to coherent and incoherent parts of the emitted energy, showing that the comparison of observations in and out of the forward direction may provide an additional hint for the relevance of nonlinear dynamical effects in the prediction of Fe${}^{16+}$ line-strength ratios.

\subsubsection{Off the forward direction}

\begin{figure*}[t]
\includegraphics[clip=true, width=.49 \textwidth]{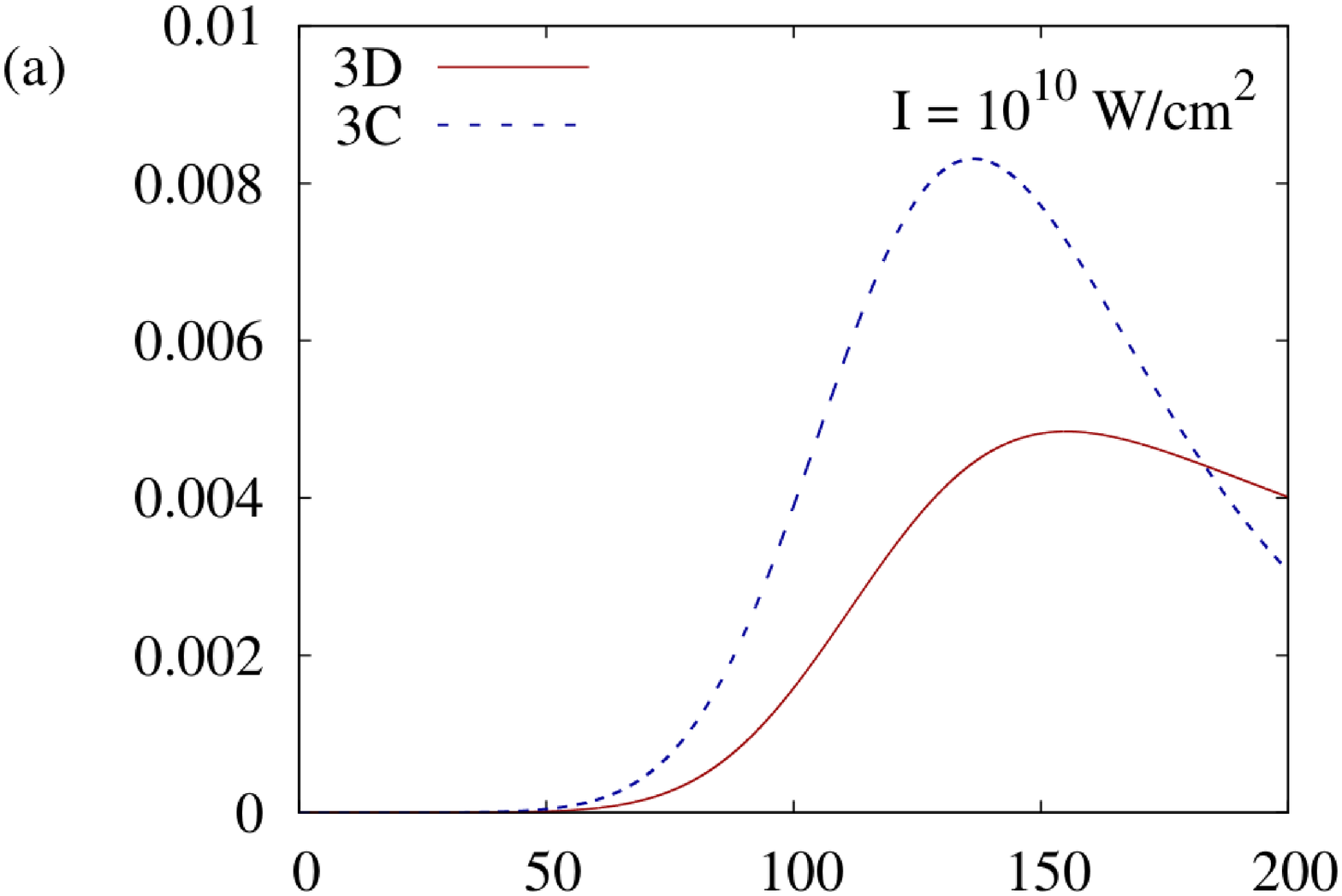}
\includegraphics[clip=true, width=.49 \textwidth]{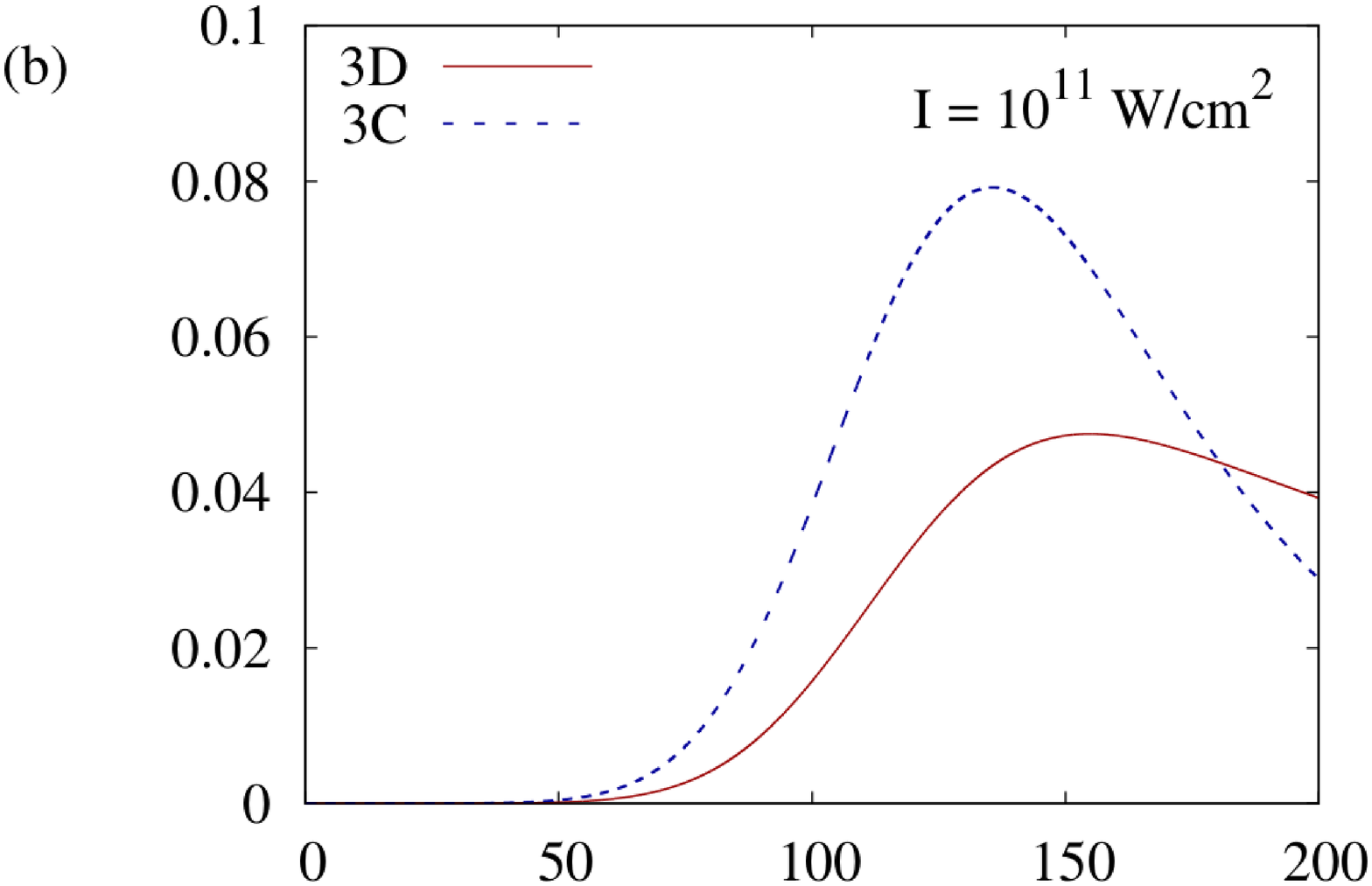}
\includegraphics[clip=true, width=.49 \textwidth]{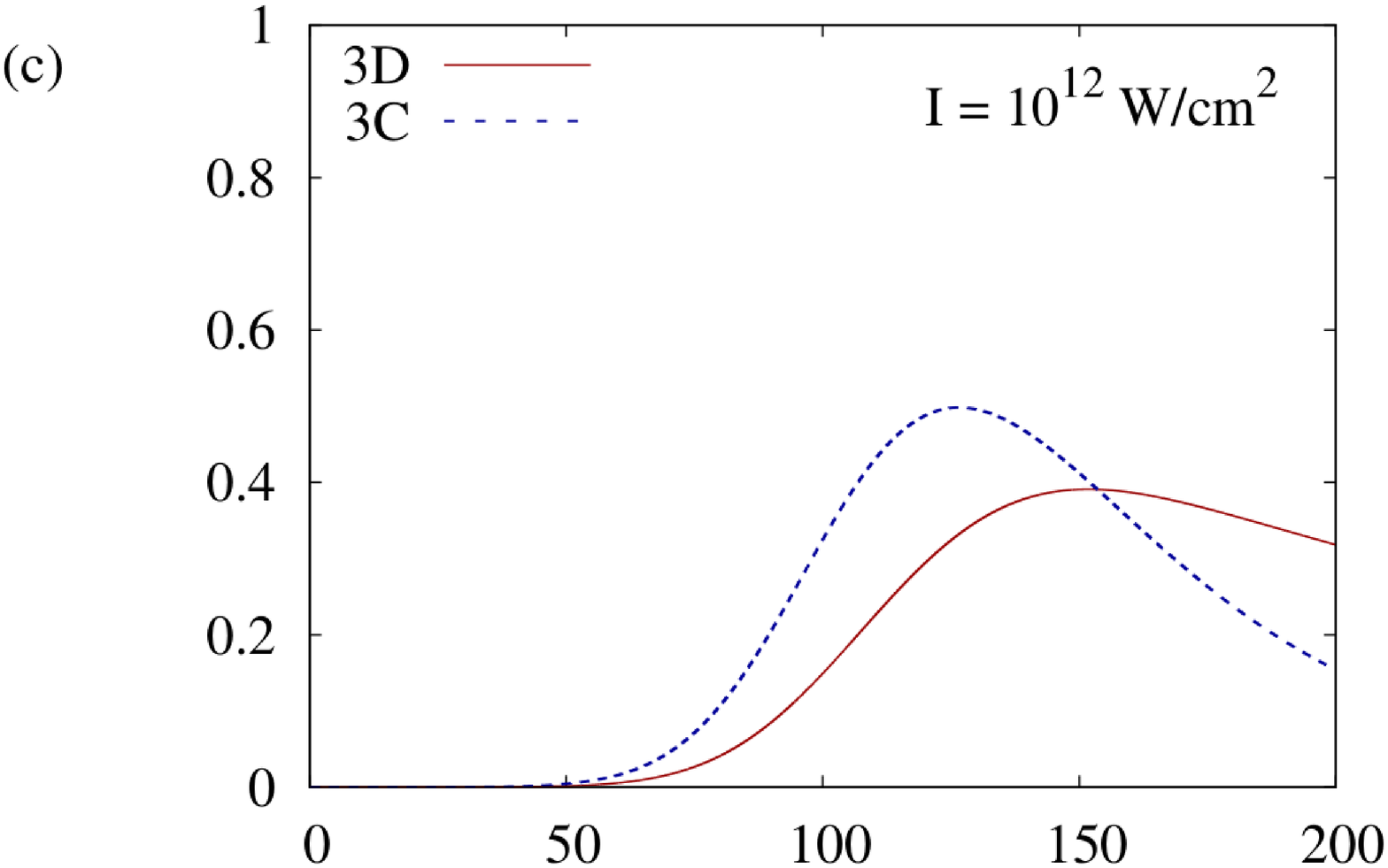}
\includegraphics[clip=true, width=.49 \textwidth]{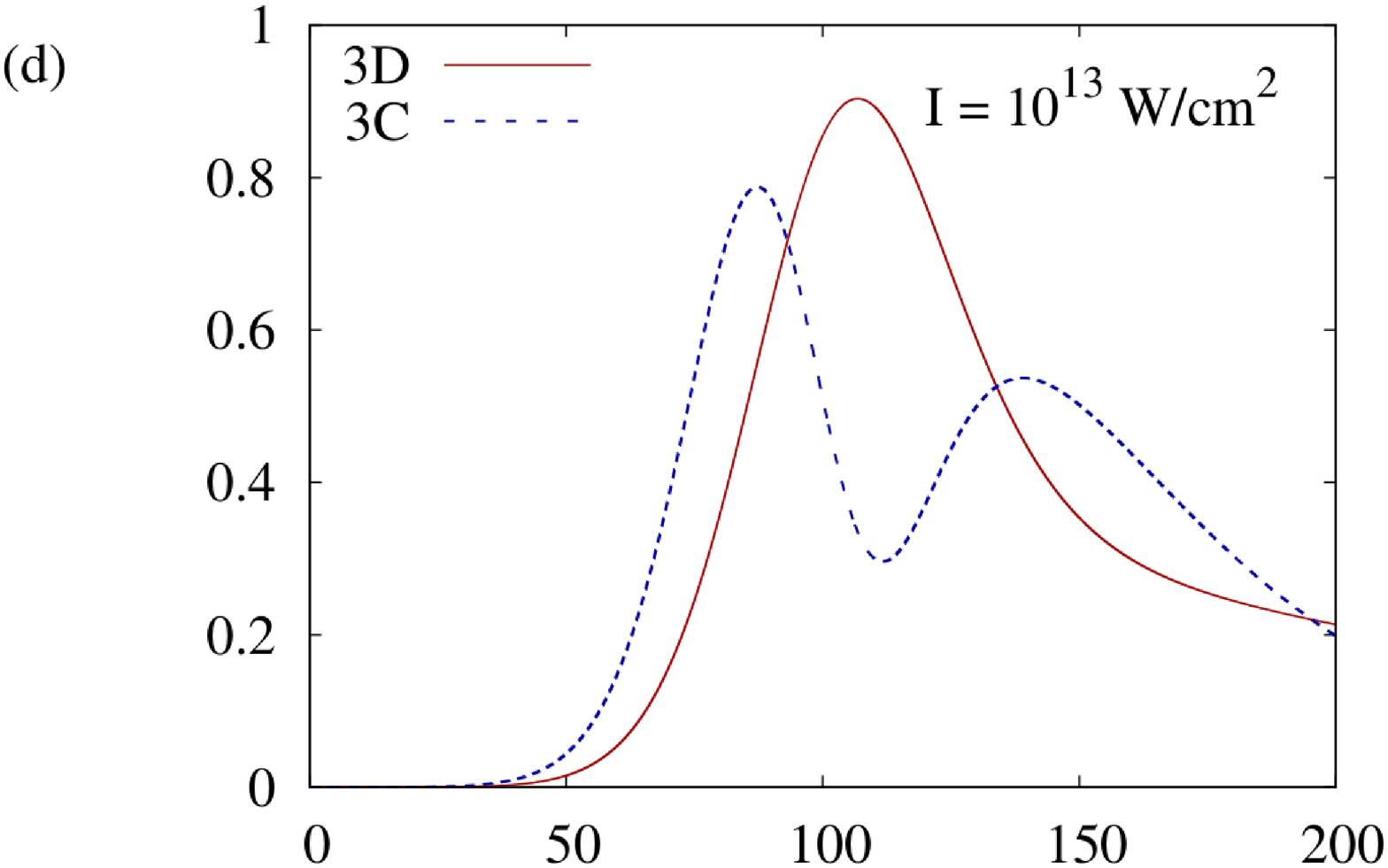}
\caption{(color online) Population of the excited $2p^53d$ states for the 3C and 3D lines, as a function of time. 
The curves are given for a short pulse duration of 200~fs and different pulse intensities, and display the time evolution of the system in the presence of the pulse.}
\label{fig:population}
\end{figure*}

In Fig.~\ref{fig:population}, we show the time evolution of the population of the excited 3C and 3D states in the presence of a Gaussian pulse of duration $T=200$~fs and for different intensities. 
This corresponds to typical LCLS parameters, with nominal intensities in the range of $I=10^{11}-10^{14}$~W/cm$^2$ and pulse durations 
in the range of $T=200-2000$~fs \cite{LCLS,Bernitt2012}.
For Gaussian pulses we use the pulse envelope $E(t) = E_{\rm max} e^{- \frac{t^2}{T^2} 32 \ln(2)}$ and a constant phase $\psi(t)=0$.
In Figs.~\ref{fig:population}(a) and \ref{fig:population}(b), for $I=10^{10}$~W/cm${}^2$ and $I=10^{11}$~W/cm${}^2$ respectively,
one is still in the weak-field regime and the emitted energy is low. 
The figures exhibit almost identical overall shapes, with maximal values of population proportional to the pulse intensity and thus differing by one order of magnitude.
This agrees with the linear theory of resonance fluorescence.
For an intensity of $I=10^{12}$~W/cm${}^2$, displayed in Fig.~\ref{fig:population}(c), the shape of the curve differs from the previous ones and a decrease in the 3C/3D ratio starts to be visible. 
For an intensity of $I=10^{13}$~W/cm${}^2$, shown in Fig.~\ref{fig:population}(d), Rabi oscillations appear, implying a high sensitivity of the 3C/3D ratio to the pulse parameters.

\begin{figure*}[t]
\includegraphics[clip=true, width=.9 \textwidth]{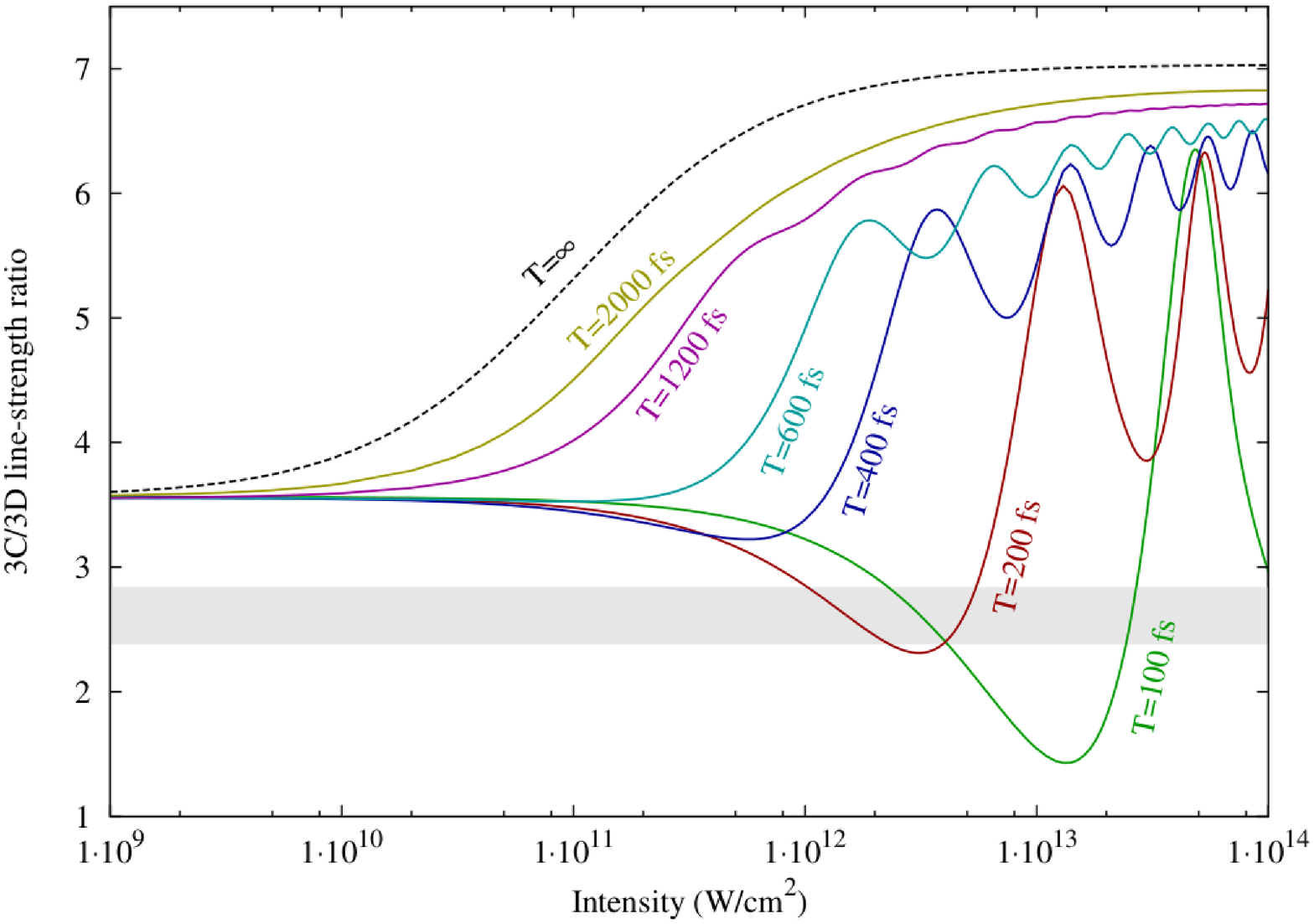}
\caption{(color online) The line-strength ratio $S_{\rm 3C}/S_{\rm 3D}$ as a function of the intensity and duration of the Gaussian pulse. 
The analytical result for a continuous-wave field is shown with a dashed line.
The gray shaded area shows the experimental ratio 2.61(23)~\cite{Bernitt2012}.}
\label{fig:function}
\end{figure*}

The 3C/3D ratio as a function of pulse intensity is plotted in Fig.~\ref{fig:function} for Gaussian pulses of different durations.
At low intensities, a value of 3.56 is approached for every pulse duration, in agreement with the {linear (weak-field)} theory.
However, at increasing intensities different nonlinear features are exhibited for different pulse durations. 
In particular, at decreasing values of the pulse duration, oscillations of the line-strength ratio are shown for high intensities. 
Shorter pulses are associated with larger oscillations in the line-strength ratio, implying a larger sensitivity to the intensity.
For the longest pulse duration considered, $T=2000$~fs, there are no visible oscillations in the ratio and the behavior approaches the one predicted for a continuous-wave field, given by Eq.~\eqref{eq:analyticalratio}.
Furthermore, we notice that, approximately between $I=1-5\times 10^{12}$ W/cm$^2$ and for $T=200$~fs, the resulting line-strength ratio agrees with the measured value of 2.61$\pm$0.23~\cite{Bernitt2012}.

These results confirm the importance of strong-field dynamical effects in a relatively intense XFEL field, suggesting that the linear model may not have been appropriate for the interpretation of the experimental results from Ref.~\cite{Bernitt2012}. 
We point out that the linear model is applicable for intensities low compared to the dipole-moment-dependent saturation intensity $I_{\rm sat}$. This implies that, for smaller values of the dipole-moment matrix elements, nonlinear dynamical effects would appear at larger intensities.

\begin{figure*}[t]
\includegraphics[clip=true, width=0.9\textwidth]{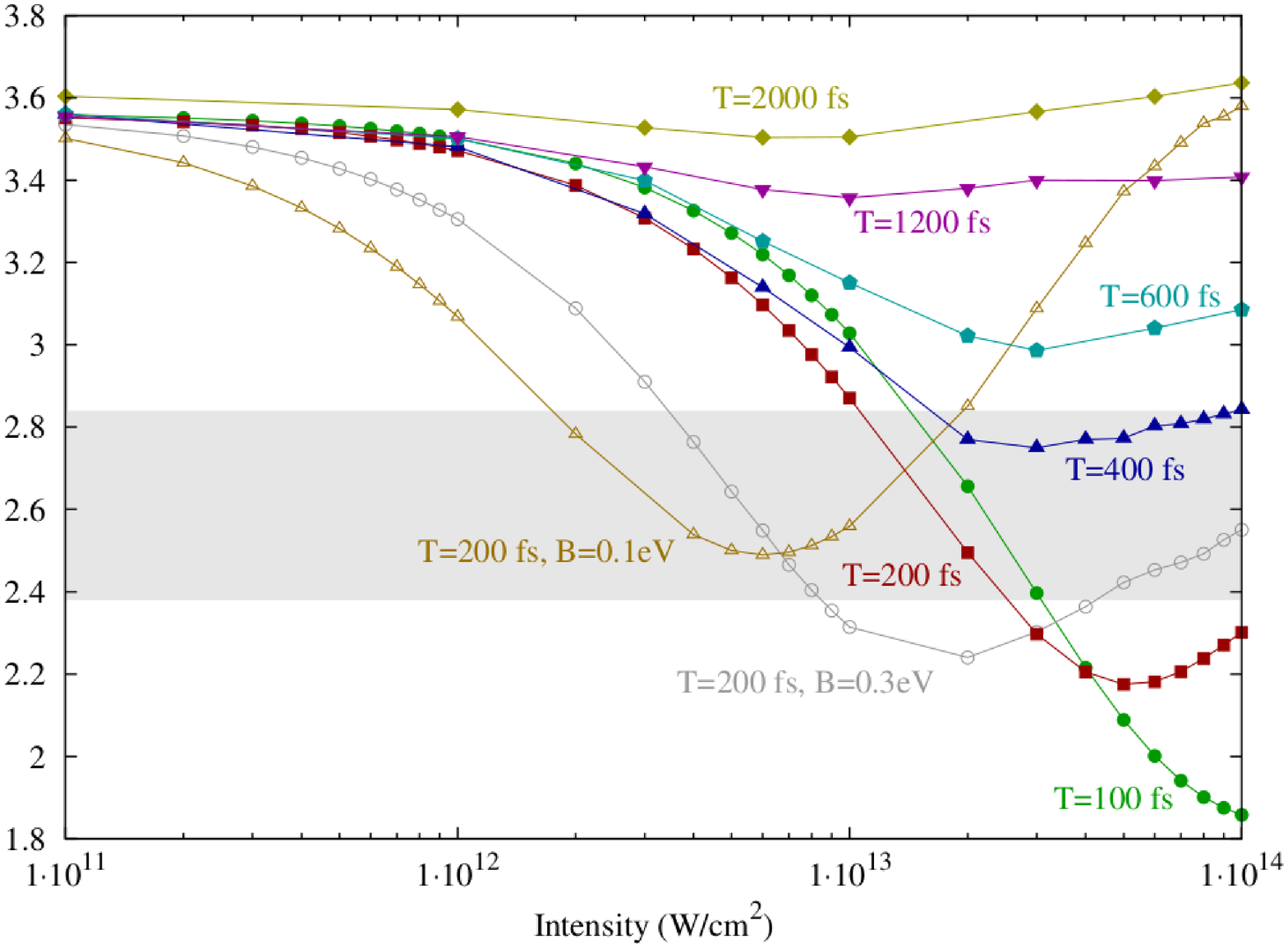}
\caption{(color online) The line-strength ratio
$S_{\rm 3C}/S_{\rm 3D}$ as a function of the intensity $I$ and duration $T$ of the chaotic pulse with a bandwidth of $B=1$~eV.
The gray shaded area shows the experimental ratio 2.61(23)~\cite{Bernitt2012}.}
\label{fig:avpulse}
\end{figure*}

Because of the  chaotic nature of XFEL pulses, Gaussian pulses are not sufficient for an adequate simulation of the true experimental scenario.
Experimental XFEL pulses generated via self-amplified spontaneous emission \cite{Bonifacio1984} are here modeled via the partial-coherence method from Refs.~\cite{Pfeifer2010,Cavaletto2012}.
In the simulations, we employ a series of such randomized pulses to integrate the equations of motion \eqref{eq:mastereq}.
The results of our simulations are shown in Fig. \ref{fig:avpulse} for different pulse durations and intensities.
The bandwidth of the pulses was chosen to be 1~eV, which approximately corresponds to the experimentally observed line width.
For each point in Fig.~\ref{fig:avpulse}, an average over 10 independent random chaotic pulses was taken.
The simulation uncertainty can be estimated to be on the $1-2\%$ level.
As in the case of Gaussian pulses, the line-strength ratio clearly depends on the pulse parameters. Also here, for small intensities, the value of 3.56 is approached, in agreement with the weak-field limit.
We notice that, for chaotic pulses, the effect of the decrease in the 3C/3D line-strength ratio can be observed within a wider range of pulse intensities than for Gaussian pulses, as well as for a significantly larger interval of pulse durations.
However, in contrast to the case of driving Gaussian pulses, no visible increase in the ratio towards the limit of continuous-wave fields from Eq.~(\ref{eq:analyticalratio}) can be distinguished. 
Due to the fixed bandwidth of $B = 1$~eV, all chaotic pulses present a series of short spikes with a coherence time of $\approx \hbar/B$.
As a result, line-strength ratios calculated for long chaotic pulses are much lower than those calculated for correspondingly long Gaussian pulses.
This is mainly due to the short coherence time, such that the effect of a long chaotic pulse should be rather compared to the results obtained with many uncorrelated, short and weak Gaussian pulses. 

\begin{figure*}[t]
\includegraphics[clip=true, width=0.49 \textwidth]{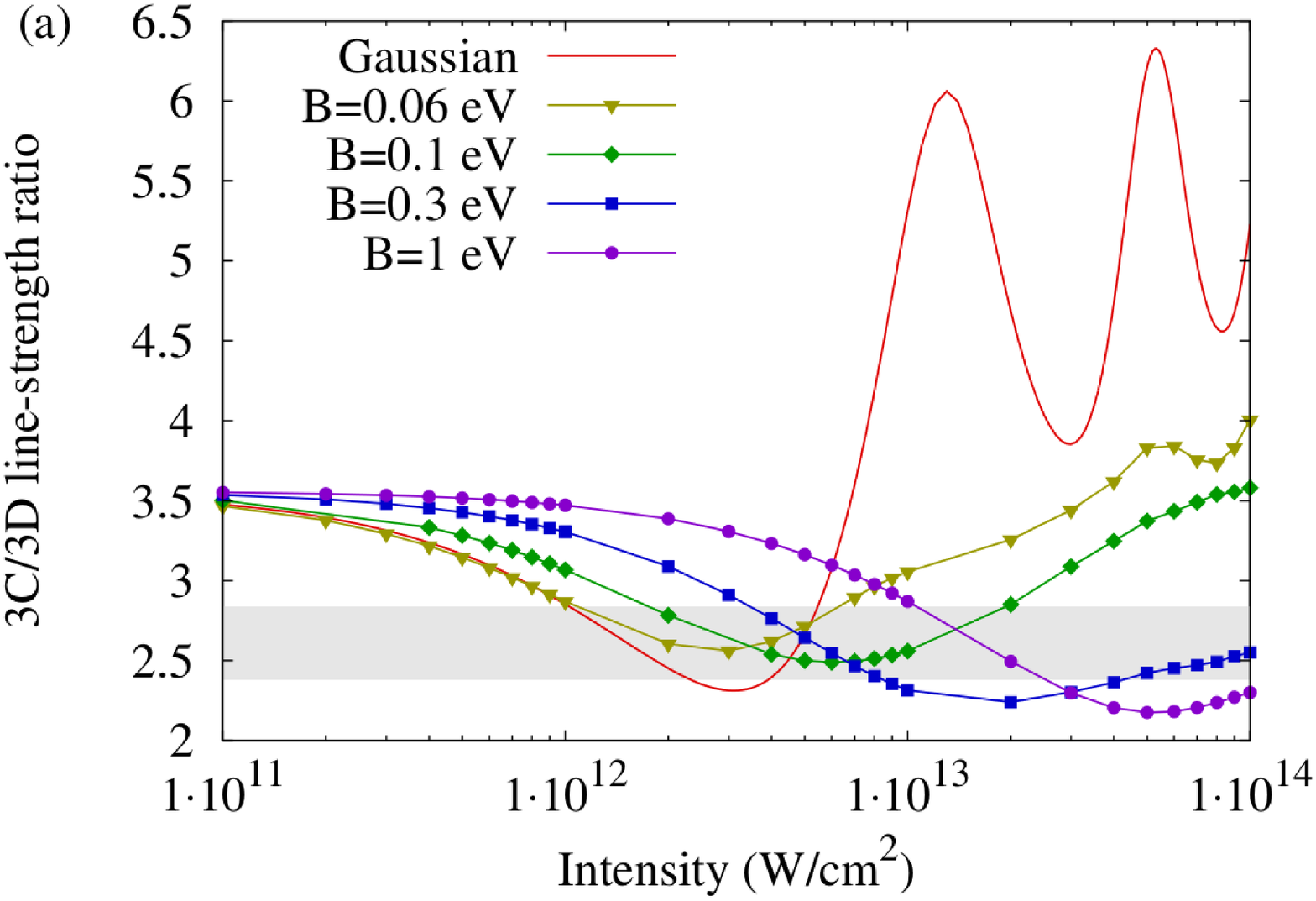}
\includegraphics[clip=true, width=0.49 \textwidth]{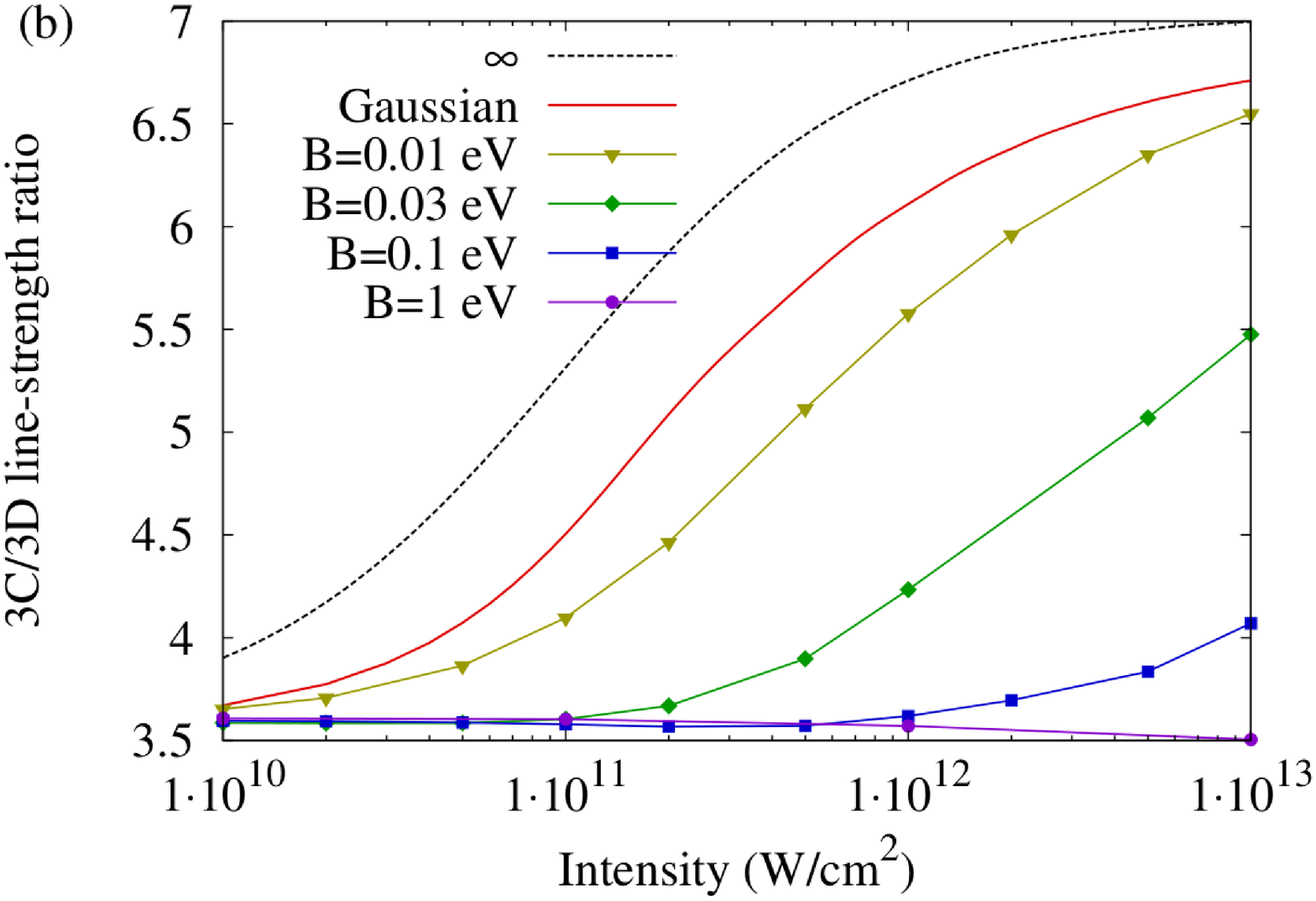}
\caption{(color online) The line-strength ratio
$S_{\rm 3C}/S_{\rm 3D}$ as a function of the intensity $I$ and bandwidth $B$ for (a) short  $T = 200$~fs and (b) long  $T=2000$~fs pulses.
In plot (b), the analytical result for a continuous-wave field is shown with a dashed line.
The gray shaded area shows experimental results as usual.}
\label{fig:bandwidth}
\end{figure*}

In order to understand the effect of the XFEL pulse bandwidth, in Fig.~\ref{fig:bandwidth} we present results for (a) short $T=200$~fs, and (b) long $T=2000$~fs chaotic pulses with different bandwidths, comparing them with results obtained from Gaussian pulses with the same duration. 
For the long pulses in Fig.~\ref{fig:bandwidth}(b), the continuous-wave-field limit~\eqref{eq:analyticalratio} is also plotted.
The results show that incoherence effects become more significant at increasing values of the bandwidth of the chaotic pulse, i.e., for larger FWHM of the energy spectrum.
The decrease in bandwidth (increase in coherence time) corresponds to a more coherent pulse, approaching a behavior closer to that displayed
by fully coherent, i.e. transform-limited Gaussian pulses.
In the experiment, a monochromator was used to obtain pulses with better coherence, i.e. smaller bandwidth, which also leads to a decrease in  pulse intensities. 
However, our simulations show that for smaller bandwidths also smaller intensities are needed to reach the experimentally observed ratio of 2.61$\pm$0.23. 
{Therefore, we can conclude that effects due to the use of the monochromator can be accounted for  within our modeling.}

We finally point out that parameters of XFEL pulses are not fixed from pulse to pulse because of their chaotic nature, yet we estimate an average peak intensity lying in the range of $10^{11} - 10^{14}$~W/cm${}^2$.
The previous results were obtained by averaging over different pulses of fixed duration $T$ and intensity $I$.
In the experiment, however, the measured line-strength ratio may have originated from an average over pulses differing in $I$ and $T$.
In order to better model the experimental conditions, we calculated the line-strength ratio by averaging with equal weight over pulses of fixed duration and varying intensity. 
Thereby, we observed that the main contribution to this averaged line-strength ratio comes from pulses of larger intensity, associated with lower values of the ratio.
This is an additional point hinting to the relevance of nonlinear dynamical effects in the modeling of the experimental results.
 
\subsubsection{In the forward direction}

\begin{figure*}[t]
\includegraphics[clip=true, width=.49 \textwidth]{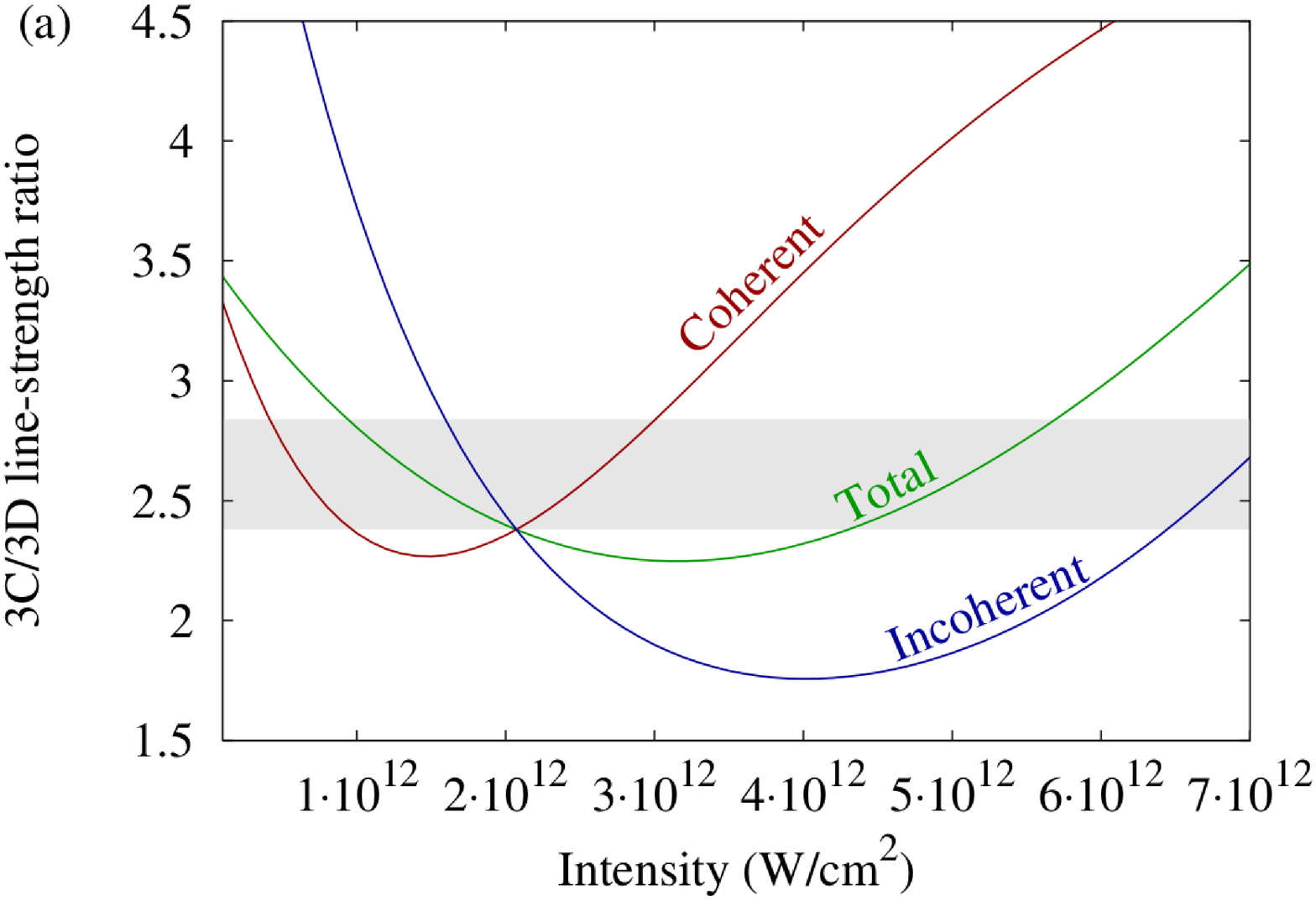}
\includegraphics[clip=true, width=.49 \textwidth]{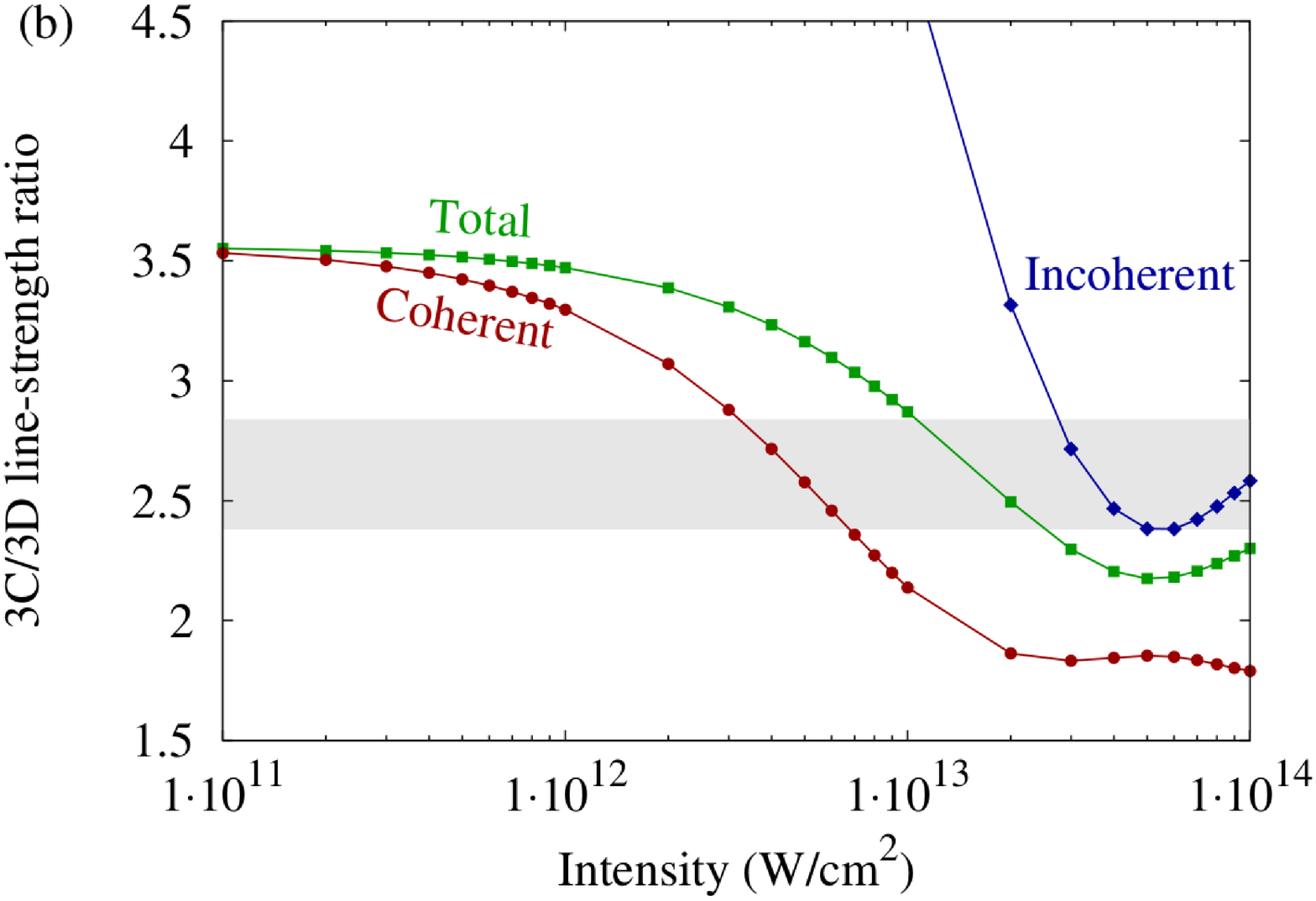}
\caption{(color online) Total, coherent and incoherent part of the line-strength ratio
$S_{\rm 3C}/S_{\rm 3D}$ as a function of the intensity $I$ for (a) Gaussian  and (b) chaotic pulses with  duration $T=200$~fs.
For chaotic pulses an average over 10 realizations with a bandwidth of $B=1$~eV was employed.
The gray shaded area shows experimental results as usual, with the measured total contribution off the forward direction.
}
\label{fig:coherent}
\end{figure*}

In order to further investigate the dependence of the above results upon the position of the x-ray detector, the different contributions of coherent, incoherent, and total emission to the line-strength ratio~\eqref{eq:spectrum} are shown in Fig.~\ref{fig:coherent}(a) for Gaussian pulses of fixed duration and varying peak intensity. 
As displayed in Fig.~\ref{fig:geometry}, out of the forward direction, where $\mathcal{E} = \mathcal{E}_{\mathrm{coh}} + \mathcal{E}_{\mathrm{inc}}$, both coherent and incoherent parts need be included. 
In this case, as already known from the theory of the spectrum of resonance fluorescence~\cite{Jentschura20041, Scully}, 
$\mathcal{E}_{\mathrm{coh}}/\mathcal{E}\rightarrow 1$ for low peak intensities, while $\mathcal{E}_{\mathrm{coh}}/\mathcal{E}\rightarrow 0$ 
at high peak intensities. 
This is confirmed by Fig.~\ref{fig:coherent}(a), where the line-strength ratios including only $\mathcal{E}_{\mathrm{coh}}$ approach 
the total-line-strength ratios for low intensities, while they diverge for high intensities. 
As previously discussed, however, in the forward direction one has $\mathcal{E} \approx \mathcal{E}_{\mathrm{coh}}$: the emitted energy due to the coherent part of the spectrum of resonance fluorescence dominates and determines the measured line-strength ratio. 
We notice that, at sufficiently high intensities, all three curves displayed in Fig.~\ref{fig:coherent}(a) enter the shaded area, corresponding to line-strength ratios measured in the experiment. 
Interestingly, line-strength ratios calculated from $\mathcal{E}_{\mathrm{coh}}$ approach the experimental values for lower peak intensities than when computed from the total emitted energy $\mathcal{E}_{\mathrm{coh}} + \mathcal{E}_{\mathrm{inc}}$. 
A similar behavior is exhibited by Fig.~\ref{fig:coherent}(b) for 10 realizations of chaotic pulses of fixed duration and bandwidth and varying intensity.

The contribution of the incoherent energy $\mathcal{E}_{\mathrm{inc}}$ to the total energy increases with the intensity of the x-ray pulse. 
By comparing measurement results in the forward direction, with the only contribution of the coherent energy $\mathcal{E}_{\mathrm{coh}}$, 
and in a different direction, where $\mathcal{E}_{\mathrm{coh}} + \mathcal{E}_{\mathrm{inc}}$ is detected, differences in the line-strength 
ratios should be apparent, as predicted in Fig.~\ref{fig:coherent} as a consequence of nonlinear dynamical effects. This additional indication 
of the presence of strong-field effects could be easily verified experimentally.

\subsection{Influence of the uncertainty of atomic-structure calculations }
\label{Influence of the atomic calculations uncertainty}

\begin{figure*}[t]
\includegraphics[clip=true, width=.49 \textwidth]{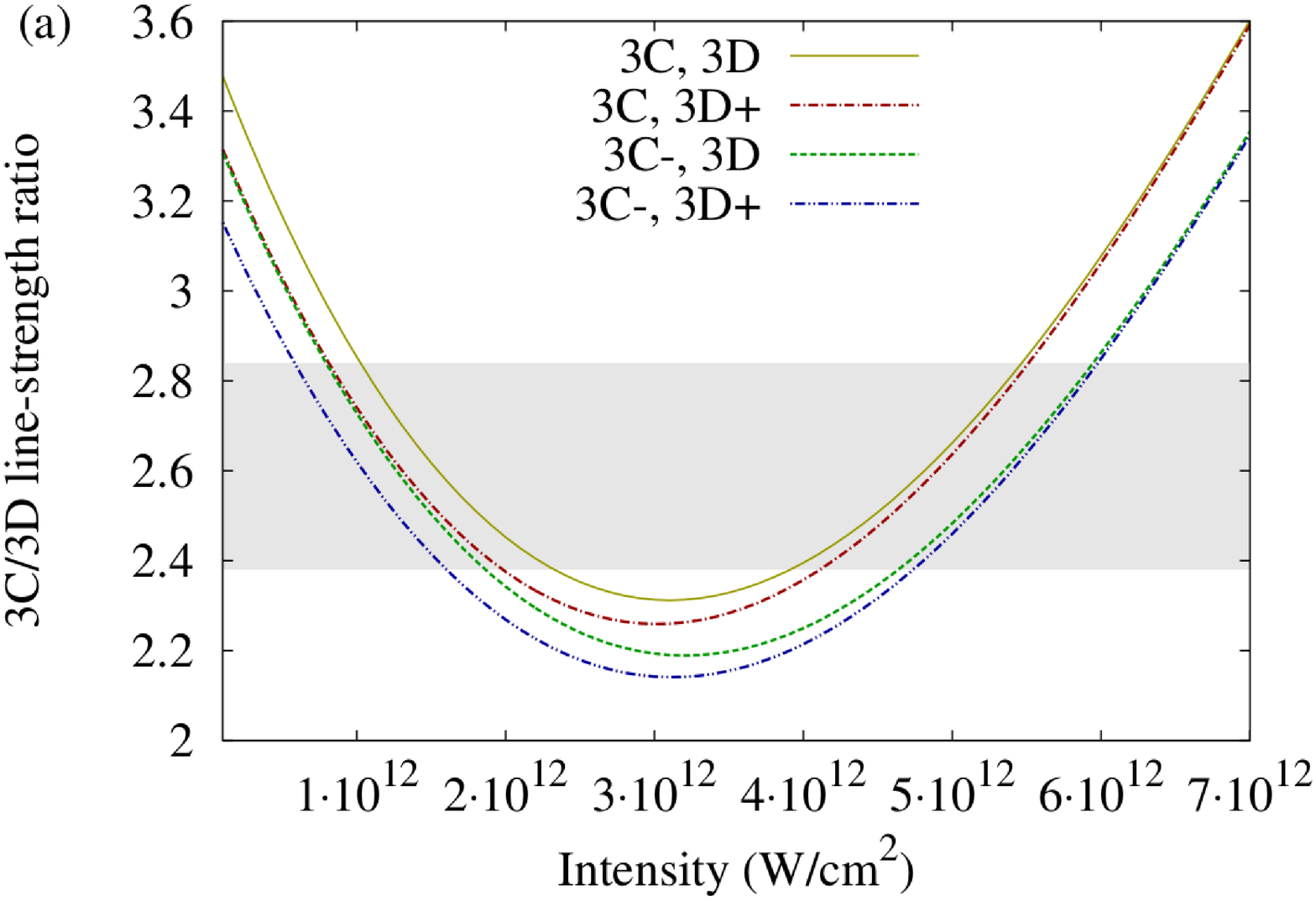}
\includegraphics[clip=true, width=.49 \textwidth]{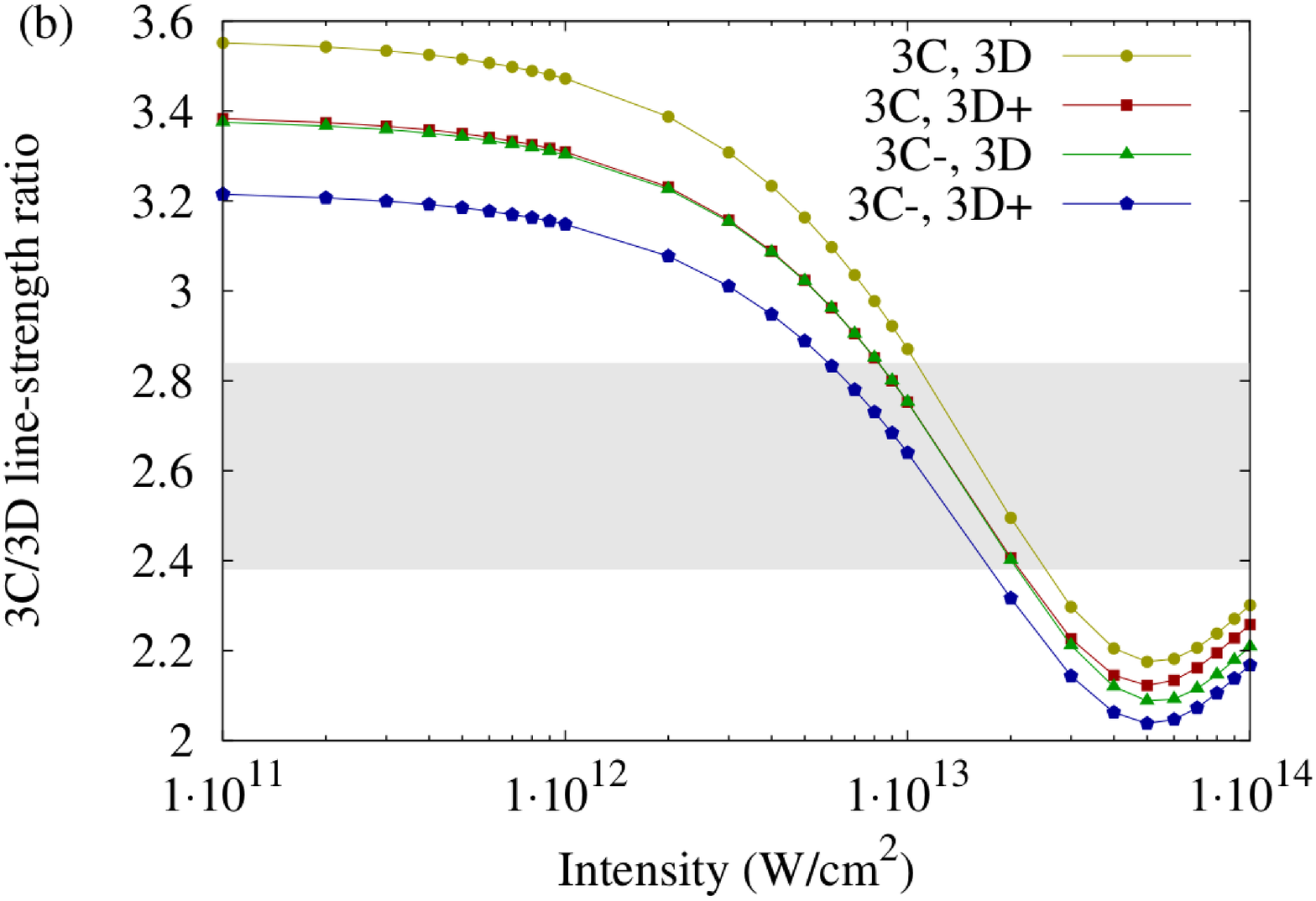}
\caption{(color online) Influence of the theoretical uncertainty of calculated Einstein coefficients to the line-strength ratio
$S_{\rm 3C}/S_{\rm 3D}$ as a function of the intensity $I$ for (a) Gaussian and (b) chaotic pulses with duration $T=200$~fs.
The minus and plus signs in the legend stand for the Einstein coefficients to be 5\% less or more, correspondingly, than our CI-DFS prediction.
For chaotic pulses an average over 10 realizations with bandwidth $B=1$~eV was employed.
The gray shaded area shows experimental results as usual.
}
\label{fig:5percent}
\end{figure*}

Theoretical predictions for the Einstein coefficients of the 3C and 3D lines and, therefore, for the weak-field value of the 
ratio under consideration $S_{\rm 3C}/S_{\rm 3D}$, are generally in agreement. 
However, small differences in the results obtained from distinct atomic many-body calculations were reported. 
Our CI-DFS method predicts a value of 3.56, and other recent theoretical predictions are in the range 3.43 -- 3.66 \cite{Bernitt2012, PhysRevA.64.012507, PhysRevA.76.062708, NIST_ASD, PhysRevA.91.012502, Godefroid_priv, Santana2015}.
We plot, for Gaussian pulses in Fig.~\ref{fig:5percent}(a) and for chaotic pulses in Fig.~\ref{fig:5percent}(b), intensity-dependent line-strength ratios for different values of the atomic parameters, such as Einstein coefficients.
By either decreasing $A_{\rm 3C}$ by 5\%, or by increasing $A_{\rm 3D}$ by 5\% with respect to their CI-DFS predicted values, 
the $S_{\rm 3C}/S_{\rm 3D}$  ratio decreases by 5\%, whereas the minimal intensity needed to reach the  value measured experimentallly decreases by 20\%, from $1.1\times 10^{13}$ to $8\times 10^{12}$~W/cm${}^2$. 
When both $A_{\rm 3C}$ and $A_{\rm 3D}$ are modified, $S_{\rm 3C}/S_{\rm 3D}$ decreases by 10\% and the minimal intensity decreases by 40\% to $6\times 10^{12}$~W/cm${}^2$. 
This shows that an uncertainty in atomic-structure parameters would increase the acceptable range of intensities for which our calculations predict results in agreement with the experiment of Ref.~\cite{Bernitt2012}.


\subsection{Analysis of the Fe${}^{15+}$ spectrum}
\label{Influence of Fe15+ lines}
 
In the experiment of Ref.~\cite{Bernitt2012} not only Fe${}^{16+}$, but also Fe${}^{15+}$ ions (and also other charge states which are not relevant in the current discussion) were present in the electron-beam ion trap. 
After measuring a spectrum of Fe${}^{15+}$ only, the electron acceleration voltage of the trap was increased and Fe${}^{16+}$ ions were also produced. 
The exact amount of each type of ion in the trap is not known exactly. 
Then, a new spectrum with both types of Fe ions  was recorded. 
In order to have a Fe${}^{16+}$ spectrum only, the normalized Fe${}^{15+}$ spectrum was subtracted from the spectrum of both ions.

We analyzed the behavior of the Fe${}^{15+}$ spectrum as a function of the intensity. 
In this case, in addition to spontaneous decay, Auger decay is possible. 
Spontaneous-decay rates $\Gamma_{eg}$ calculated with the CI-DFS method and Auger-decay rates $\Gamma_A$ calculated with the multiconfiguration Dirac-Fock (MCDF) method \cite{Bernitt2012,grasp2K,Harman2006, Zimmerer1990} are shown in Table~\ref{tab:fe15}. 
The calculated rates are in good agreement with experimental resonant photoionization spectra~\cite{Beilmann}.
The most relevant lines here are line B at 815~eV and line C at 812~eV.
The line C of Fe${}^{15+}$ is mixed with the 3D line of Fe${}^{16+}$ in a spectrum of both ions, therefore, it should be substracted.
The line B is the strongest line of both spectra and was used for normalization.

\begin{table}
\begin{tabular}{l l r r r}
\hline\hline
Line	& Transition	& \hspace{-1cm}Energy, eV	& \quad $\Gamma_{eg}$, $1/$s	& \quad $\Gamma_A$, $1/$s	\\
\hline
C	& $(2p^63s)_{\frac{1}{2}} \rightarrow (2p^53s3d)_{\frac{3}{2}}$	& 812	& $1.41\cdot10^{13}$	& $2.1\cdot10^{13}$	\\
B	& $(2p^63s)_{\frac{1}{2}} \rightarrow (2p^53s3d)_{\frac{1}{2}}$	& 815	& $2.19\cdot10^{13}$	& $1.7\cdot10^{10}$	\\
A	& $(2p^63s)_{\frac{1}{2}} \rightarrow (2p^53s3d)_{\frac{3}{2}}$	& 820	& $6.80\cdot10^{12}$	& $4.0\cdot10^{13}$	\\
\hline\hline
\end{tabular}
\caption{Transitions of Fe${}^{15+}$, with their notations, energies, spontaneous-decay rates $\Gamma_{eg}$ calculated with the CIDFS method and Auger-decay rates $\Gamma_A$ calculated with the MCDF method. } \label{tab:fe15}
\end{table}

\begin{figure}[t]
\includegraphics[clip=true, width=.49 \textwidth]{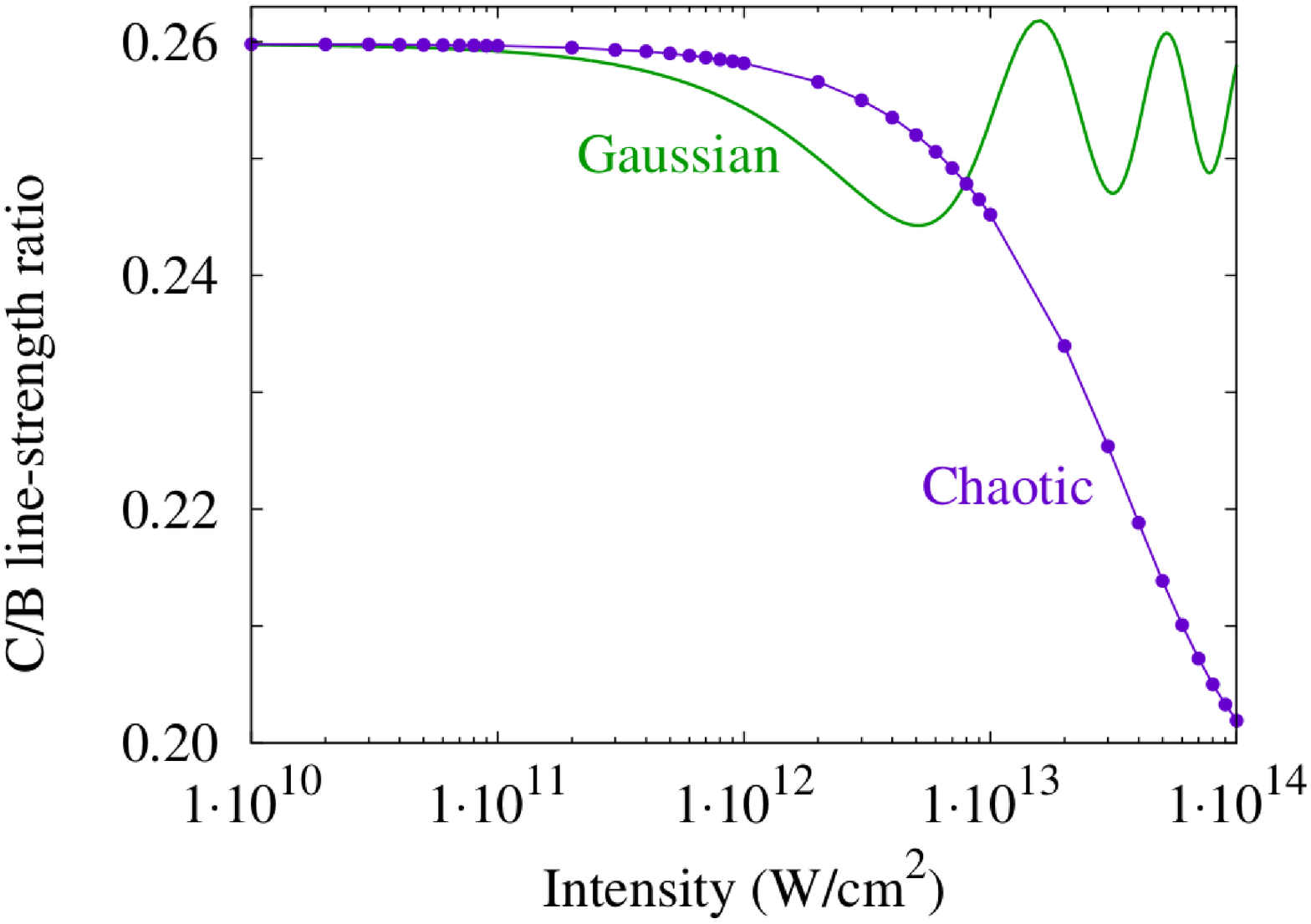}
\caption{(color online) 
Simulated intensity of Fe${}^{15+}$ C line, normalized by the strongest Fe${}^{15+}$ B line, as a function of the intensity for Gaussian and chaotic pulses with duration $T=200$~fs and bandwidth $B=1$~eV for chaotic pulses. {The geometry of the detector is assumed to be as in the experiment \cite{Bernitt2012}, see Appendix~\ref{Fe15+ line-strength ratios} for  details.} 
}
\label{fig:fe15}
\end{figure}

\begin{figure}[t]
\includegraphics[clip=true, width=.49 \textwidth]{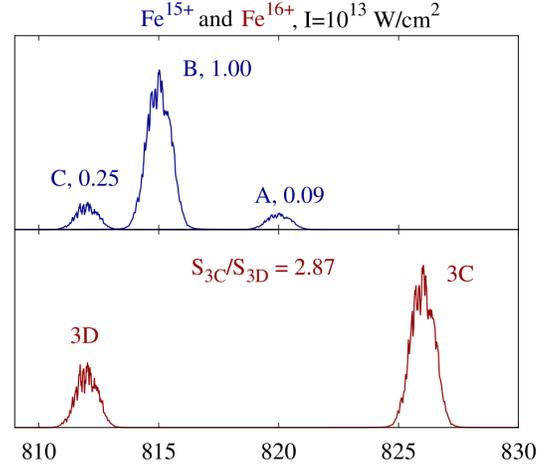}
\caption{(color online) 
Spectrum of Fe${}^{15+}$ (blue, top) and Fe${}^{16+}$ (red, bottom) for  intensity $I=10^{13}$~W/cm${}^2$, duration $T=200$~fs and bandwidth $B=1$~eV, averaged over 10 independent realizations of the pulse.
}
\label{fig:fe15_16}
\end{figure}

The parameters of Fe${}^{15+}$ transitions are close to the parameters of Fe${}^{16+}$. 
As a result, the saturation intensities $I_{\rm sat}$ are on the same order of magnitude, i.e., $10^{11}$~W/cm${}^2$, and a similar sensitivity of the line-strength ratio to the interaction parameters is expected.
In Fig.~\ref{fig:fe15} the C/B line-strength ratio as a function of intensity is presented.
In Fig.~\ref{fig:fe15_16} spectra of both ions are displayed for an intensity of $1\times 10^{13}$ W/cm${}^2$. 

{As apparent from Figs.~\ref{fig:function}, \ref{fig:avpulse} and \ref{fig:fe15}, both spectra are sensitive to the intensity, though the sensitivity of the Fe${}^{15+}$ spectrum is lower. 
{In the wide range $I=10^{10}-10^{14}$~W/cm${}^2$ the C/B ratio changes by only 20\% and, within the experimental error bars, it can be regarded as constant, as found in Ref.~\cite{Bernitt2012}, see Supplementary Figure S3 therein.}
Therefore, an analogously wide range of intensities in the case of Fe${}^{16+}$ can not be excluded. }

\section{Summary}

In summary, we conclude that dynamical processes can modify the x-ray fluorescence spectra of highly charged {Fe${}^{15+}$} and Fe${}^{16+}$ ions
when driven with strong and short x-ray free-electron-laser pulses. 
Oscillator strengths were calculated with a large-scale configuration-interaction Dirac-Fock-Sturm method including single, double and triple excitations, confirming previous theoretical calculations and their discrepancy with the experiment. 
QED corrections to transition energies were estimated with an effective screening potential and found to be negligible. 
A strong dependence of the observed 3C/3D ratio on experimental parameters, such as XFEL pulse intensity, duration and bandwidth was stated. 
By decreasing atomic calculations results of the $S_{\rm 3C}/S_{\rm 3D}$ ratio by 5\%, we decrease the minimal intensity needed to reach the experimental measured value by 20\%.
We also find that the line-strength ratio of the C and B lines in Fe${}^{15+}$ is much less sensitive to the XFEL intensities.
As a result, a wide interval of the intensities may be considered as possible for Fe${}^{16+}$ as well.  
Furthermore, we propose that comparing the energy emitted in the forward direction and in any other direction can give a hint for the limitations in the applicability of the weak-field atomic theory.

Concluding, we would like to stress that an accurate control of the pulse parameters such as intensity, duration and bandwidth is required for a proper analysis of the experimental data. 
Furthermore, dynamical effects may be relevant for both laboratory measurements and for astrophysical observations with naturally incoherent x-ray sources.
Approaching the saturation intensity, the weak-field atomic theory employed so far is shown not to be applicable anymore, which is equally valid for corresponding astrophysical scenarios. 
For instance, in the accretion disk of stellar-mass black holes in x-ray binaries~\cite{Shapiro1983}, the Eddington luminosity of $10^{18}$~W/cm$^2$ at a distance close to the innermost stable orbit of three Schwarzschild radii leads to an intensity of approximately $3\times 10^{13}$~W/cm$^2$  for a line at 1 keV with a 1-eV width, requiring the inclusion of the previously neglected nonlinear dynamics.


\section{Acknowledgements}

We acknowledge helpful advice and computer codes from Ilya I. Tupitsyn and insightful conversations with 
Sven Bernitt, Jos\'e R. Crespo L\'opez-Urrutia, Alberto Benedetti and J\"org Evers.

\appendix

\section{Two-level-model approach}
\label{Two-level model}
\begin{figure}[t]
\includegraphics[clip=true, width= 0.8\linewidth, keepaspectratio]{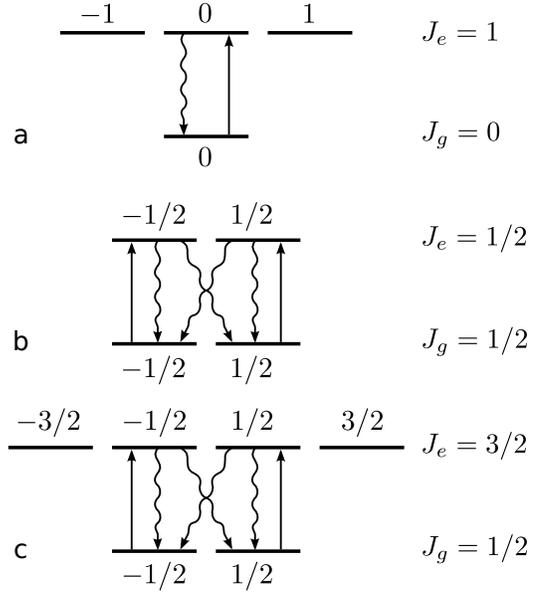}
\caption{Level schemes for the x-ray transitions responsible for (a) the 3C and 3D lines in Fe${}^{16+}$; (b) the B line in Fe${}^{15+}$; and (c) the A and C lines in Fe${}^{15+}$.}
\label{fig:levelschemes}
\end{figure}

The level schemes of the atomic transitions considered throughout the paper are displayed in Fig.~\ref{fig:levelschemes}. 
We denote with $|eM_e\rangle$ one of the degenerate excited states associated with energy $\omega_e$ and total angular-momentum quantum numbers $J_e$ and its projection $M_e$. 
We similarly write $|gM_g\rangle$ for one of the degenerate ground states of energy $\omega_{g}$ and angular-momentum quantum numbers $J_g$ and $M_g$. 
We describe the atomic system via the density matrix 
\begin{equation}
\hat{\rho}(t) = \sum_{i,\,j\in\{e,\,g\}}\sum_{M_i,M_j} \rho_{iM_i;jM_j} |iM_i\rangle \langle jM_j|,
\end{equation}
whose time evolution is given by the master equation~(\ref{eq:mastereq}).

For an external electric field $\vec{\mathscr{E}}(t)=E(t)\cos(\omega_{\mathrm{X}}t+\psi(t))\,\hat{\boldsymbol{e}}_z$, linearly polarized along the $z$ quantization axis, only transitions 
with the same quantum number $M$ are driven, as displayed by the straight arrows in Fig.~\ref{fig:levelschemes}. 
In the rotating-wave approximation, the resulting light-ion-interaction Hamiltonian is given by
\begin{equation}
\begin{split}
\hat{H}_{\mathrm{int}} = \sum_{M} &\,-\frac{\hbar\Omega_{R,M}(t)}{2}\,e^{i\omega_{\mathrm{X}}t}|gM\rangle\langle eM| \\ 
&\,-\frac{\hbar\Omega_{R,M}^*(t)}{2}\,e^{-i\omega_{\mathrm{X}}t}|eM\rangle\langle gM|, 
\end{split}
\end{equation}
with the $M$-dependent Rabi frequency
\begin{equation}
\Omega_{R,M}(t) = \frac{e}{\hbar}E(t)\langle gM| \hat{z} | eM \rangle e^{i\psi(t)}.
\end{equation}
Here, $\langle gM| e\hat{z} | eM \rangle$ is the dipole-moment matrix element, with $\hat{z}$ being the position operator along the quantization axis. 
For the atomic systems displayed in Fig.~\ref{fig:levelschemes} 
$|\langle gM| e\hat{z} | eM \rangle|$ does not depend on $M$.

Spontaneous decay from the excited to the ground level is described by the Liouville superoperator
\begin{equation}
\begin{split}
\mathcal{L}\hat{\rho}(t) = &\,\sum_{M_e, M_g} - \frac{\Gamma_{eM_e;gM_g}}{2}\biggl[ |eM_e\rangle\langle eM_e|\hat{\rho}(t)\\
&\,- |gM_g\rangle\langle eM_e|\hat{\rho}(t) |eM_e\rangle\langle gM_g| \biggr] + \mathrm{H.\,c.}
\end{split}
\end{equation}
Here, $\Gamma_{eM_e;gM_g}$ is the E1 decay rate from the excited state $|eM_e\rangle$ to the ground state $|gM_g\rangle$, 
\begin{equation}
\Gamma_{eM_e;gM_g} = \frac{4\alpha\omega_{eg}^3 |\langle gM_g |\hat{\boldsymbol{r}}| eM_e\rangle|^2}{3 c^2},
\label{eq:definitionGammaeMgM}
\end{equation}
with $\langle gM_g |\hat{\boldsymbol{r}}| eM_e\rangle$ being the electric-dipole-moment matrix element written in terms of the position operator $\hat{\boldsymbol{r}}$. 
We denote with $\hat{r}_M$ 
the three standard spherical components of the first-order irreducible tensor operator $\hat{\boldsymbol{r}}$. 
The Wigner-Eckart theorem allows one to write the dipole-moment matrix element $\langle gM_g |\hat{r}_M| eM_e\rangle$ as the product of the Clebsch-Gordan coefficient $C_{J_gM_g1M}^{J_eM_e}$  times a term which is independent of the projections, i.e., 
\begin{equation}
\langle gM_g |\hat{r}_M| eM_e\rangle = \frac{\langle gJ_g ||\hat{r}^{(1)}||eJ_e\rangle}{\sqrt{2J_g+1}} C_{J_eM_e1M}^{J_gM_g} 
\end{equation}
where $\langle gJ_g ||\hat{r}^{(1)}||eJ_e\rangle$ stands for the reduced matrix element. 
As a result, 
\begin{equation}
\Gamma_{eM_e;gM_g} \sim \frac{|\langle gJ_g ||\hat{r}^{(1)}||eJ_e\rangle|^2}{2J_g+1} (C_{J_eM_e1(M_g-M_e)}^{J_gM_g})^2. 
\label{eq:GammaeMgM}
\end{equation}

By exploiting the general properties of the Clebsch-Gordan coefficients,
one can conclude that the decay rate 
\begin{equation}
\Gamma_{eM_e;g} = \sum_{M_g} \Gamma_{eM_e;gM_g},
\label{eq:GammaeMg}
\end{equation}
from the excited state $|eM_e\rangle$ to the ground level, is independent of the quantum number $M_e$ of the initial excited state, namely,
\begin{equation} \label{eq:formulalunga}
\Gamma_{eM_e;g}\sim \frac{|\langle gJ_g ||\hat{r}^{(1)}||eJ_e\rangle |^2}{2J_e + 1},
\end{equation}
and it hence coincides with the total decay rate
\begin{equation}
\Gamma_{eg} = \frac{1}{2J_e + 1}\sum_{M_e} \Gamma_{eM_e;g} = \Gamma_{eM_e;g}
\label{eq:Gammaeg}
\end{equation}
from the excited level to the ground level. Furthermore, by comparing Eqs.~(\ref{eq:GammaeMgM}), (\ref{eq:formulalunga}) and (\ref{eq:Gammaeg}), it follows that
\begin{equation}
\Gamma_{eM_e;gM_g} = \Gamma_{eg} \frac{2J_e+1}{2J_g+1} (C_{J_eM_e1(M_g-M_e)}^{J_gM_g})^2. 
\label{eq:relationGammas}
\end{equation}

Autoionization from the excited level results in the emission of a free electron and the generation of a bound system with lower charge. We denote with $|fM_f\rangle$ this total final state, with given angular-momentum quantum numbers $J_f$ and $M_f$, implicitly assuming a sum over the quantum numbers of the free ionized electron. In the master equation~(\ref{eq:mastereq}) this is modeled by
\begin{equation}
\mathcal{D}\hat{\rho}(t) = \sum_{M_e} - \frac{\Gamma_{\mathrm{A},M_e}}{2} |eM_e\rangle\langle eM_e|\hat{\rho}(t) + \mathrm{H.\,c.},
\end{equation}
describing an effective loss of population from the bound excited state $|eM_e\rangle$ at the Auger-decay rate
\begin{equation}
\Gamma_{\mathrm{A},M_e} \sim \sum_{M_f} |\langle fM_f |\hat{V} | eM_e \rangle|^2,
\end{equation}
with $\hat{V}$ being the electron-electron Coulomb-interaction operator. 
In the cases under consideration, the leading term of the operator $\hat{V}$ has rank 0. 
This implies that the Auger-decay rate $\Gamma_{\mathrm{A},M_e} $ is independent of $M_e$ and, therefore, identical to the total Auger-decay rate
\begin{equation}
\Gamma_{\mathrm{A}} = \frac{1}{2J_e+1}\sum_{M_e}\Gamma_{\mathrm{A},M_e} = \Gamma_{\mathrm{A},M_e}.
\label{eq:GammaAM}
\end{equation}

We write down the master equation~(\ref{eq:mastereq}) for the different elements of the density matrix:
\begin{equation}
\begin{split}
\frac{d \rho_{gM;gM'}}{dt} = &+\frac{i}{2}\Omega_{R,M}(t)e^{i\omega_{\mathrm{X}}t} \rho_{eM;gM'} \\
&- \frac{i}{2}\Omega^*_{R,M'}(t)e^{-i\omega_{\mathrm{X}}t} \rho_{gM;eM'} \\
&+ \delta_{MM'}\sum_{M''}\Gamma_{eM'';gM}\rho_{eM'';eM''};
\end{split}
\label{eq:Agg}
\end{equation}
\begin{equation}
\begin{split}
\frac{d \rho_{eM;eM'}}{dt} = &-\frac{i}{2}\Omega_{R,M'}(t)e^{i\omega_{\mathrm{X}}t} \rho_{eM;gM'} \\
&+ \frac{i}{2}\Omega^*_{R,M}(t)e^{-i\omega_{\mathrm{X}}t} \rho_{gM;eM'} \\
&-\frac{1}{2}\sum_{M''}(\Gamma_{eM;gM''} + \Gamma_{eM';gM''})\rho_{eM;eM'}\\
&-\frac{1}{2}(\Gamma_{\mathrm{A},M} + \Gamma_{\mathrm{A},M'})\rho_{eM;eM'};
\end{split}
\label{eq:Aee}
\end{equation}
\begin{equation}
\begin{split}
\frac{d \rho_{eM;gM'}}{dt} = &-i\omega_{eg} \rho_{eM;gM'}\\
&-\frac{i}{2}\Omega^*_{R,M'}(t)e^{-i\omega_{\mathrm{X}}t} (\rho_{eM;eM'} - \rho_{gM;gM'}) \\
&- \frac{1}{2}\bigl(\Gamma_{\mathrm{A},M} + \sum_{M''}\Gamma_{eM;gM''} \bigr)\rho_{eM;gM'};
\end{split}
\label{eq:Aeg}
\end{equation}
and $\rho_{gM';eM} = \rho_{eM;eM'}^*$. We assume that the atomic system is initially in the ground level, such that each one of the $(2J_g+1)$ ground states $|g, M_g\rangle$ is occupied with equal probability $1/(2J_g+1)$. We also assume that the initial coherences -- also among ground states differing in the angular-momentum quantum number $M$ -- vanish. By closely inspecting Eqs.~(\ref{eq:Agg}), (\ref{eq:Aee}), and (\ref{eq:Aeg}), one can notice that, with such natural initial conditions, 
if $M\neq M'$ then $\rho_{i,M;j,M'}(t) = 0$ at any time. Among all the differential equations from Eqs.~(\ref{eq:Agg}), (\ref{eq:Aee}), and (\ref{eq:Aeg}), we can therefore focus exclusively on $(2J+1)$ sets of 4 differential equations, each set describing the evolution of the $M$th transition, $|M| \leq J = \min(J_e,J_g)$. 

By bearing in mind Eqs.~(\ref{eq:GammaeMg}), (\ref{eq:Gammaeg}), and (\ref{eq:GammaAM}), we can write Eqs.~(\ref{eq:Agg}), (\ref{eq:Aee}), and (\ref{eq:Aeg}) as follows:
\begin{equation}
\begin{split}
\frac{d \rho_{gM;gM}}{dt} = &+\frac{i}{2}\Omega_{R,M}(t)e^{i\omega_{\mathrm{X}}t} \rho_{eM;gM} \\
&- \frac{i}{2}\Omega^*_{R,M}(t)e^{-i\omega_{\mathrm{X}}t} \rho_{gM;eM} \\
&+ \sum_{M'}\Gamma_{eM';gM}\rho_{eM';eM'};
\end{split}
\label{eq:Mprimedependent}
\end{equation}
\begin{equation}
\begin{split}
\frac{d \rho_{eM;eM}}{dt} = &-\frac{i}{2}\Omega_{R,M}(t)e^{i\omega_{\mathrm{X}}t} \rho_{eM;gM} \\
&+ \frac{i}{2}\Omega^*_{R,M}(t)e^{-i\omega_{\mathrm{X}}t} \rho_{gM;eM} \\
&-(\Gamma_{eg}+\Gamma_{\mathrm{A}}) \rho_{eM;eM};
\end{split}
\label{eq:Aeebis}
\end{equation}
\begin{equation}
\begin{split}
\frac{d \rho_{eM;gM}}{dt} = &-i\omega_{eg} \rho_{eM;gM}\\
&-\frac{i}{2}\Omega^*_{R,M}(t)e^{-i\omega_{\mathrm{X}}t} (\rho_{eM;eM} - \rho_{gM;gM}) \\
&- \frac{1}{2}(\Gamma_{\mathrm{A}} + \Gamma_{eg} )\rho_{eM;gM}.
\end{split}
\label{eq:Aegbis}
\end{equation}
We notice that the term describing spontaneous decay from all $(2J+1)$ excited states to the $M$th ground state is the only term connecting the dynamics of the $M$th transitions to those of the remaining $2J$ ones. 
Yet, for the atomic systems displayed in Fig.~\ref{fig:levelschemes}, the Rabi frequencies $\Omega_{R,M}(t)$ are (up to a minus sign) independent of $M$. 
Along with the $M$-independent initial conditions $\rho_{iM;jM}(0) = \delta_{i,g}\delta_{jg}/(2J_g+1)$, one can easily conclude that $\rho_{eM;eM}(t) = \rho_{eM';eM'}(t)$, for any $M$, $M'$ and for any time. 
Furthermore, for the three atomic systems depicted in Fig.~\ref{fig:levelschemes}, Eq.~(\ref{eq:relationGammas}) implies that $\sum_{M'} \Gamma_{eM';gM}=\Gamma_{eg}$. As a result, Eq.~(\ref{eq:Mprimedependent}) reads
\begin{equation}
\begin{split}
\frac{d \rho_{gM;gM}}{dt} = &+\frac{i}{2}\Omega_{R,M}(t)e^{i\omega_{\mathrm{X}}t} \rho_{eM;gM} \\
&- \frac{i}{2}\Omega^*_{R,M}(t)e^{-i\omega_{\mathrm{X}}t} \rho_{gM;eM} \\
&+ \Gamma_{eg}\rho_{eM;eM},
\end{split}
\end{equation}
leading to $(2J+1)$ identical, independent sets of differential equations, each one describing the two-level system associated with the $M$th transition. This justifies the two-level approximation presented in Subsection~\ref{Modeling of ion--x-ray interaction}.

\section{Simulation of the observed Fe${}^{15+}$ line-strength ratios}
\label{Fe15+ line-strength ratios}

In contrast to the 3C and 3D lines in Fe${}^{16+}$ ions -- both originating from the $0 \rightarrow 1$ transition displayed in Fig.~\ref{fig:levelschemes}(a) --, the A, B, and C lines in Fe${}^{15+}$ result from transitions with different angular-momentum quantum numbers $J$. 
As shown in Table~\ref{tab:fe15}, the A and C lines result from a $3/2\rightarrow 1/2$ transition, see Fig.~\ref{fig:levelschemes}(c), while the B line from a $1/2 \rightarrow 1/2$ transition, see Fig.~\ref{fig:levelschemes}(b). 
Hence, the angular distribution of the emitted radiation is not the same for the three transitions and Fe${}^{15+}$ x-ray line-strength ratios are affected by the position of the detector.

In the following, we derive the formulas used for the numerical calculations presented in Subsection~\ref{Influence of Fe15+ lines}. 
We assume the geometry depicted in Fig.~\ref{fig:geometry} and a detector positioned along the $z$-axis, in accordance with the experimental measurement from Ref.~\cite{Bernitt2012}. 
In order to calculate the detected emitted energy, we first relate the electric-field operator to the dipole response of the excited atomic system. 
In the far-zone approximation and in terms of the ladder operators $\hat{\sigma}_{iM_i;jM_j} = |iM_i\rangle \langle jM_j |$, the positive-frequency electric-field operator at time $t$ and along the $\hat{\boldsymbol{e}}_z$ unit vector is proportional to
\begin{align} \label{eq:electricfieldoperator}
& \hat{\boldsymbol{E}}^{+}(\hat{\boldsymbol{e}}_z, t) \sim \{\hat{\boldsymbol{r}}(t) - [\hat{\boldsymbol{r}}(t)\cdot \hat{\boldsymbol{e}}_z]\hat{\boldsymbol{e}}_z\}\,\omega_{eg}^2 = \\
&\sum_{M_g, M_e}\sum_{M\in\{-1,1\}} \hat{\sigma}_{gM_g;eM_e}(t) \langle gM_g |\hat{r}_{M}| eM_e\rangle\,\hat{\boldsymbol{e}}_{M} \,\omega_{eg}^2. \notag
\end{align}
Here, $\hat{\boldsymbol{r}} = \sum \hat{r}_M\,\hat{\boldsymbol{e}}_M$ is written in terms of its three standard spherical components $\hat{r}_M$
and complex unit vectors $\hat{\boldsymbol{e}}_M$.
The emitted energy $\mathcal{E}(\hat{\boldsymbol{e}}_z)$ detected along the $z$ direction is related to the electric-field operator $\hat{\boldsymbol{E}}^{+}(\hat{\boldsymbol{e}}_z, t)$ via
\begin{equation}
\mathcal{E}(\hat{\boldsymbol{e}}_z) = \int_{-\infty}^{\infty}\langle \hat{\boldsymbol{E}}^{-}(\hat{\boldsymbol{e}}_z, t)\cdot \hat{\boldsymbol{E}}^{+}(\hat{\boldsymbol{e}}_z, t)\rangle\, dt,
\end{equation}
where $\hat{\boldsymbol{E}}^{-}(\hat{\boldsymbol{e}}_z, t) = [\hat{\boldsymbol{E}}^{+}(\hat{\boldsymbol{e}}_z, t)]^{\dagger}$ and $\langle \ldots\rangle$ stands for the expectation value of the operator. 
By employing Eq.~(\ref{eq:electricfieldoperator}) along with the orthogonality of the  unit vectors $\hat{\boldsymbol{e}}_M$, this leads to
\begin{align}
\mathcal{E}(\hat{\boldsymbol{e}}_z) \sim &\sum_{\substack{M_g, M_g', \\ M_e, M_e'}}\sum_{\substack{M\in \\ \{-1,1\}}} 
\langle gM_g |\hat{r}^{\dagger}_{M}| eM_e\rangle \langle eM_e' |\hat{r}_{M}| gM_g'\rangle \omega_{eg}^4 \notag \\
& \times\int_{-\infty}^{\infty}\langle \hat{\sigma}_{eM_e;gM_g}(t)\, \hat{\sigma}_{gM_g';eM_e'}(t)\rangle\,dt.
\end{align}
As discussed in Appendix~\ref{Two-level model}, the expectation values
\begin{equation}
\begin{split}
\langle \hat{\sigma}_{eM_e;gM_g}(t) & \hat{\sigma}_{gM_g';eM_e'}(t)\rangle \\
= &\mathrm{Tr}\{|eM_e\rangle\langle gM_g| gM_g'\rangle\langle eM_e'|\hat{\rho}(t)\} \\
= &\delta_{M_gM_g'}\,\rho_{eM_e';eM_e}(t)
\end{split}
\end{equation}
differ from zero 
only if $M_e = M_e'= \tilde{M}$, with $ |\tilde{M}| \leq J= \min(J_g,J_e)$. 
As a result, the emitted energy $\mathcal{E}(\hat{\boldsymbol{e}}_z)$ reduces to
\begin{equation}
\begin{split}
\mathcal{E}(\hat{\boldsymbol{e}}_z) \sim&\sum_{M_g, \tilde{M}}\sum_{M\in\{-1,1\}} |\langle e\tilde{M} |\hat{r}_{M}| gM_g\rangle|^2 \\
&\ \ \ \times\,\omega_{eg}^4\,\int_{-\infty}^{\infty} \rho_{e\tilde{M};e\tilde{M}}(t) \,dt,
\end{split}
\label{eq:emenergyFe15+}
\end{equation}
where $\rho_{e\tilde{M};e\tilde{M}}(t)$ does not depend explicitly on $\tilde{M}$, as discussed in Appendix~\ref{Two-level model}.

As displayed in Figs.~\ref{fig:levelschemes}(b) and \ref{fig:levelschemes}(c), for the A, B, and C lines in Fe${}^{15+}$ ions, Eq.~(\ref{eq:emenergyFe15+}) reduces to
\begin{align}
\mathcal{E}(\hat{\boldsymbol{e}}_z) \sim&\,\biggl(|\langle g\frac{1}{2} |\hat{r}_1 | e\frac{-1}{2}\rangle|^2  + |\langle g\frac{-1}{2} |\hat{r}_{-1} | e\frac{1}{2}\rangle|^2\biggr) \notag \\
& \times\,\omega_{eg}^4\,\int_{-\infty}^{\infty} \rho_{e\tilde{M};e\tilde{M}}(t) \,dt \label{eq:emenergyFe15+_final} \\
\sim &\,(\Gamma_{e\frac{-1}{2};g\frac{1}{2}}  + \Gamma_{e\frac{1}{2};g\frac{-1}{2}}) \omega_{eg} \notag
\int_{-\infty}^{\infty} \rho_{e\tilde{M};e\tilde{M}}(t) dt,
\end{align}
where we have used Eqs.~(\ref{eq:definitionGammaeMgM}). Further use of Eq.~(\ref{eq:relationGammas}) allows one to write Eq.~(\ref{eq:emenergyFe15+_final}) as
\begin{equation}
\begin{split}
\mathcal{E}(\hat{\boldsymbol{e}}_z) \sim&\,\frac{2J_e+1}{2J_g+1} 
\biggl[ (C^{J_g1/2}_{J_e- 1/211})^2 + (C^{J_g-1/2}_{J_e1/21-1})^2\biggr] \\
&\ \ \ \times\, \Gamma_{eg}\,\omega_{eg}\,\int_{-\infty}^{\infty} \rho_{e\tilde{M};e\tilde{M}}(t) \,dt.
\end{split}
\label{eq:lastformula}
\end{equation}
This implies that $\mathcal{E}(\hat{\boldsymbol{e}}_z) $ is explicitly dependent on the angular-momentum quantum numbers $J_e$ and $J_g$ of the excited and ground levels, respectively. 
Hence, the multiplication factor 
$$\frac{2J_e+1}{2J_g+1} \biggl[ (C^{J_g1/2}_{J_e- 1/211})^2 + (C^{J_g-1/2}_{J_e1/21-1})^2\biggr]$$
need be included when calculating strength ratios of lines with different $J_e$. 
For the B line ($1/2 \rightarrow 1/2$) this factor equals $4/3$, while for the A and C lines ($3/2 \rightarrow 1/2$) it is equal to $2/3$, so one should include a factor of $1/2$ to the C/B line-strength  ratio to simulate the experimentally observed one \cite{Bernitt2012} in the far-zone approximation. 

In contrast to the case described above, the 3C and 3D lines in Fe${}^{16+}$ are both associated with transitions $0 \rightarrow 1$, as depicted in Fig.~\ref{fig:levelschemes}(a). Line-strength ratios are therefore not dependent upon the position of the detector and are not affected by the $J$-dependent factors appearing in Eq.~(\ref{eq:lastformula}). This justifies the use of Eqs.~(\ref{eq:E-Delta}) and (\ref{eq:spectrum}) for Fe${}^{16+}$ ions, as discussed in Subsection~\ref{Line-strength-ratio calculation}.

%

\end{document}